\documentclass[aps,prb,reprint,superscriptaddress,floatfix]{revtex4-2}

\usepackage[greek,english]{babel}
\usepackage[version=4]{mhchem}
\usepackage{graphicx}
\usepackage{siunitx}
\usepackage{textcomp}
\usepackage[colorlinks=true, allcolors=blue]{hyperref}
\usepackage{paralist}
\usepackage{orcidlink}
\usepackage{microtype}

\DeclareSIUnit[quantity-product=\,]{\GPaEXP}{GPa^{EXP}}
\DeclareSIUnit[quantity-product=\,]{\GPaQHA}{GPa^{QHA}}
\DeclareSIUnit[quantity-product=\,]{\GPaPBE}{GPa^{PBE}}
\DeclareSIUnit\angstrom{\text {Å}}

\begin{document}


\title{\textit{\textit{Ab initio}} investigation of H-bond disordering in \textgreek{δ}-AlOOH}


\author{Chenxing Luo\,\orcidlink{0000-0003-4116-6851}}
\affiliation{Department of Applied Physics and Applied Mathematics, Columbia University, New York, NY 10027, USA}

\author{Koichiro Umemoto\,\orcidlink{0000-0003-2410-4042}}
\affiliation{Earth-Life Science Institute, Tokyo Institute of Technology, Tokyo, Japan}
\affiliation{Theoretical Quantum Physics Laboratory, Cluster for Pioneering Research, RIKEN, Wako-shi, Saitama 351-0198, Japan}

\author{Renata M. Wentzcovitch\,\orcidlink{0000-0001-5663-9426}}
\email[]{rmw2150@columbia.edu}
\affiliation{Department of Applied Physics and Applied Mathematics, Columbia University, New York, NY 10027, USA}
\affiliation{Department of Earth and Environmental Sciences, Columbia University, New York, NY 10027, USA}
\affiliation{Lamont–Doherty Earth Observatory, Columbia University, Palisades, NY 10964, USA}


\date{\today}

\begin{abstract}
\textgreek{δ}-AlOOH (\textgreek{δ}) is a high-pressure hydrous phase that participates in the deep geological water cycle. At 0~GPa, \textgreek{δ} has asymmetric hydrogen bonds (H-bonds). Under pressure, it exhibits H-bond disordering, tunneling, and finally, H-bond symmetrization at $\sim$18~GPa. This study investigates these 300~K pressure-induced state changes in \textgreek{δ} with \textit{ab initio} calculations. H-bond disordering in \textgreek{δ} was modeled using supercell multi-configuration quasiharmonic calculations. We examine: (a) energy barriers for proton jumps, (b) the pressure dependence of phonon frequencies, (c) 300~K compressibility, (d) neutron diffraction pattern anomalies, and (e) compare \textit{ab initio} bond lengths with measured ones. Such thorough and systematic comparisons indicate that: (a) proton “disorder” has a restricted meaning when applied to \textgreek{δ}. Nevertheless, H-bonds are disordered between 0 and 8~GPa, and a gradual change in H-bond configuration results in enhanced compressibility. (b) several structural and vibrational anomalies at $\sim$8~GPa are consistent with the disappearance of a particular (HOC-12) H-bond configuration and its change into another one (HOC-11*). (c) between 8–11~GPa, H-bond configuration (HOC-11*) is generally ordered, at least in short- to mid-range scale. (d) between 11.5–18~GPa, H-bond lengths approach a critical value that impedes compression, resulting in decreased compressibility. In this pressure range, especially approaching H-bond symmetrization at $\sim$18~GPa, anharmonicity and tunneling should play an essential role in the proton dynamics. Further simulations accounting for these effects are desirable to clarify the protons' state in this pressure range.
\end{abstract}


\maketitle

\section{Introduction}
\label{sec:introduction}

\textgreek{δ}-AlOOH (\textgreek{δ}) with the \ce{CaCl2}-like or distorted stishovite structure \cite{suzukiNewHydrousPhase2000} is a critical, high-pressure hydrous phase in subducted slabs in the Earth’s lower mantle. The 3D network of strong Al-O-Al ionic bonds reinforces the structure and stabilizes \textgreek{δ} in a vast pressure/temperature range. Experiments have shown that \textgreek{δ} remains stable up to core-mantle boundary (CMB) pressures, $\sim$136~GPa \cite{duanPhaseStabilityThermal2018, sanoAluminousHydrousMineral2008}, and temperatures approaching the cold-slab geotherm \cite{litasovPhaseRelationsHydrous2005}. \textgreek{δ} can form solid solutions with other isostructural phases, e.g., \ce{MgSiO4H2} phase H \cite{tsuchiyaFirstPrinciplesPrediction2013} and \textgreek{ε}-FeOOH \cite{xuSolubilityBehaviorDAlOOH2019}. The existence of \textgreek{δ} substantially extends the stability field of the solid solutions, allowing the ternary system to survive extreme pressures \cite{ohiraStabilityHydrousDphase2014} and eventually transport water to the CMB region through slab subduction to participate in the global water cycle \cite{ohtaniFateWaterTransported2018}.

Although \textgreek{δ}’s crystal structure has a $P2_1nm$ space group (No.\,30) \cite{suzukiNewHydrousPhase2000}, its H-bond arrangement remained unresolved in powder X-ray diffraction and attracted much interest. \textit{Ab initio} calculations predicted \textgreek{δ} forms asymmetric H-bonds (O—H···O) at low pressures \cite{tsuchiyaFirstPrinciplesCalculation2002}. It also indicated that H-bonds should symmetrize under pressure, forming only ionic (O—H—O) bonds at $\sim$30~GPa accompanied by an increase in elastic moduli \cite{tsuchiyaFirstPrinciplesCalculation2002}. The O—H···O configuration consists of a longer hydrogen bond (H-bond), O···H, and a shorter ionic bond, O—H. Tsuchiya et al. \cite{tsuchiyaFirstPrinciplesCalculation2002} proposed two viable hydrogen off-centered (HOC) unit cell models for \textgreek{δ}, HOC-1, and HOC-2. The relative position of two protons in the \ce{CaCl2}-like primitive cell defines these models, with two likely sites for the second proton, H1 or H2, therefore HOC-1 and HOC-2 models. In the H1 site, the second proton relates to the first one by a $2_1[100]$ screw operation, while in the H2, it relates by a $2_1[010]$. HOC-1 has lower enthalpy in static in \textit{ab initio} calculations using the Perdew-Burke-Ernzerhof (PBE) generalized gradient approximation (GGA) \cite{perdewGeneralizedGradientApproximation1996}. These calculations predict H-bond symmetrization at 30~GPa. Several experimental studies confirm this asymmetric proton distribution at low pressures \cite{kudohSpaceGroupHydrogen2004, komatsuRedeterminationHighpressureModification2006}. Still, elastic moduli stiffening signaling H-bond symmetrization is seen at 8~GPa, a much lower pressure than the \textit{ab initio} static prediction \cite{mashinoSoundVelocitiesDAlOOH2016, sano-furukawaChangeCompressibilityDAlOOH2009, simonovaStructuralStudyDAlOOH2020}. In addition, proton site occupancies were observed to change under pressure, with H1 sites favored at 0~GPa \cite{komatsuRedeterminationHighpressureModification2006}, but H2 sites also being partially occupied at higher pressures \cite{sano-furukawaDirectObservationSymmetrization2018, sano-furukawaNeutronDiffractionStudy2008}. An order-disordered H-bond transition model was introduced to explain this partial occupancy change with pressure \cite{sano-furukawaDirectObservationSymmetrization2018, sano-furukawaNeutronDiffractionStudy2008}. The stiffening in elastic moduli at 8~GPa correlates with a redistribution of protons among these sites. Although H-bonds remain asymmetric after this transition, H1 and H2 sites become equally occupied beyond 8~GPa \cite{sano-furukawaDirectObservationSymmetrization2018}. This transition is associated with the extinction of the $0kl$ neutron diffraction peaks \cite{sano-furukawaDirectObservationSymmetrization2018, kuribayashiObservationPressureinducedPhase2014} of the $P2_1nm$ structure. The structure after 8~GPa is referred to as a fully-disordered or the $Pnnm$-disordered model \cite{sano-furukawaDirectObservationSymmetrization2018, sano-furukawaNeutronDiffractionStudy2008, kuribayashiObservationPressureinducedPhase2014}. Further compression would merge the H1 and H2 sites and symmetrize H-bonds at $\sim$18~GPa \cite{sano-furukawaDirectObservationSymmetrization2018}, which might still be a relatively low pressure compared to the static \textit{ab initio} (PBE) prediction of 30~GPa \cite{tsuchiyaFirstPrinciplesCalculation2002, tsuchiyaElasticPropertiesDAlOOH2009a}.

H-bond disordering is a general phenomenon in numerous hydrous and nominally anhydrous mineral phases (NAMs) \cite{qinInitioStudyWater2018, bellWaterEarthMantle1992, kohnOrderingHydroxylDefects2002}. From a microscopic perspective, H-bond disordering implies the coexistence of multiple H-bond configurations on short-, medium-, or long-range scales. In hydrous phases, this process is usually associated with H-bond symmetrization. As previously shown in H2O-ice \cite{benoitShapesProtonsHydrogen2005} and here shown in Secs.\,\ref{sec:3A} and \ref{sec:3C} below, with increasing pressure, there are simultaneous reductions in (a) the energy barrier height between H-sites and (b) in the differences between H-bond configuration energies. Both contribute to the much-increased likelihood of disorder when $d$(OO) approaches its critical value $\sim$2.4~Å in both ice-VII \cite{benoitShapesProtonsHydrogen2005,benoitQuantumEffectsPhase1998,benoitReassigningHydrogenBondCentering2002} and \textgreek{δ} \cite{sano-furukawaDirectObservationSymmetrization2018}. The energy barrier reduction also facilitates proton mobility, especially tunneling, translating into more significant proton redistribution and less well-defined H-sites or $d$(OH). In addition, proton tunneling has been shown to occur in ice \cite{benoitShapesProtonsHydrogen2005,benoitQuantumEffectsPhase1998,benoitReassigningHydrogenBondCentering2002} and is expected to occur in \textgreek{δ} \cite{sano-furukawaDirectObservationSymmetrization2018}, although not in the partial H-bond disorder stage ($\sim$5.6~GPa) \cite{trybelAbsenceProtonTunneling2021}.

The interplay between disordering, tunneling, and H-bond symmetrization, results in a complex multi-stage phenomenon. For example, a series of \textit{\textit{ab initio}} studies \cite{benoitShapesProtonsHydrogen2005,benoitQuantumEffectsPhase1998,benoitReassigningHydrogenBondCentering2002} have identified the “molecular—ionized—centered” stages throughout the H-bond symmetrization process in the ice-VIII-VII-X system. In the case of \textgreek{δ}, although similar stages of the H-bond symmetrization transition have been confirmed by experiments \cite{sano-furukawaDirectObservationSymmetrization2018} with transition boundaries constrained to $\sim$9~GPa and $\sim$18~GPa \cite{sano-furukawaDirectObservationSymmetrization2018}, \textit{ab initio} calculations have not explored this phenomenon yet. Also, how changes in H-bond arrangements lead to the observed anomalous macroscopic properties such as stiffening in compressibility \cite{sano-furukawaChangeCompressibilityDAlOOH2009} have not yet been clarified. The pressure evolution of the \ce{OH-} infrared (IR) and Raman frequencies \cite{kagiInfraredAbsorptionSpectra2010, mashinoSoundVelocitiesDAlOOH2016} is another issue that needs quantitative interpretation by \textit{ab initio} phonon calculations.

There is no lack of \textit{ab initio} studies of \textgreek{δ} in recent years. Unfortunately, most studies focus on ordered primitive cells, i.e., HOC-1 and HOC-2 (e.g., see Refs.\,\cite{tsuchiyaElasticPropertiesDAlOOH2009a,trybelAbsenceProtonTunneling2021,pillaiFirstPrinciplesStudy2018}), favoring the HOC-1 model because of its lower static enthalpy. Addressing \textgreek{δ} with a single configuration disregards the characteristic disorder \cite{sano-furukawaDirectObservationSymmetrization2018} and its effects entirely. Disorder as the coexistence of multiple H-bond configurations can be modeled by \textit{ab initio} calculations using supercells. For example, using a $1 \times 1 \times 2$ supercell accounts for the interaction of neighboring H-bond configurations along the $c$-direction. The adoption of $1 \times 1 \times 2$ supercells \cite{tsuchiyaVibrationalPropertiesDAlOOH2008} allowed the four broad peaks of \textgreek{δ} in the OH-stretching frequency regime in the ambient pressure Raman spectrum \cite{xueCationOrderHydrogen2006, ohtaniStabilityFieldNew2001} to be successfully modeled.

The success of this simple multi-configuration (\textit{mc}) model \cite{tsuchiyaVibrationalPropertiesDAlOOH2008} is not accidental. The “fully-disordered” or “$Pnnm$-disorder model” invoked in several experimental studies \cite{sano-furukawaDirectObservationSymmetrization2018, kuribayashiObservationPressureinducedPhase2014} is not entirely accurate either. This is because H-bond disordering is restricted by specific rules, such as “ice rules” in \ce{H2O}-ice \cite{bernalTheoryWaterIonic1933}. Disorder is a complex phenomenon in stoichiometric hydrous phases with high H content because these “ice-disorder”-like rules can vary in each system. In the case of \textgreek{δ}, in a regime where proton positions are well-defined, each Al-centered octahedron has a single \ce{OH-} radical associated with it. As we will show in this paper, both O—H and O···H bonds remain in the $x, y$ plane. Together, these constraints impose H-bond ordering in two dimensions ($a$- and $b$-directions), at least medium-range order (MRO), with disorder limited to a single dimension, the $c$-direction. Therefore, an ensemble of $1 \times 1 \times 2$ supercell configurations might capture the main effects of disorder in \textgreek{δ}. The coexistence of H-bond configurations can be treated with the multi-configuration quasiharmonic approximation (\textit{mc}-QHA) \cite{qinQhaPythonPackage2019}. This approach has been used to address the free energy of ice-VII \cite{qinQhaPythonPackage2019} and has successfully described the order-disorder transition in ice-VIII to -VII in a similar H-bond disordering problem \cite{umemotoOrderDisorderPhase2010}. 

H-bond symmetrization is convoluted with a spin-transition in \ce{(Al,Fe)OOH} \cite{ohiraCompressionalBehaviorSpin2019} and with Mg-Si site disorder in \ce{MgSiO4H2} \cite{tsuchiyaCrystalStructureEquation2015}. Because of the abundant and detailed studies of \textgreek{δ}’s structure at pressures below $\sim$18~GPa, the H-bond symmetrization pressure, and the unique 1D disorder that allows direct observation of proton distribution projected onto the (001) plane, \textgreek{δ} is a convenient case to study this order-disorder phenomenon that also happens in the other related phases, \ce{(Al,Fe)OOH} and \ce{MgSiO4H2}. A better understanding of disordering, tunneling, and H-bond symmetrization in \textgreek{δ} would also allow us to more confidently address these more complex cases.

This study seeks to clarify the multi-stage disorder-tunneling-symmetrization transition process in \textgreek{δ} with \textit{ab initio} calculations. We hope to provide a more detailed atomistic interpretation of macroscopic observations. Although H-bond is itself challenging for \textit{ab initio} treatments, we would still like to understand the capability and accuracy of the standard DFT functionals for such a problem. For validation against room temperature measurements, we focus on the transition sequence at 300~K. Sec.\,\ref{sec:method} first summarizes the methods and calculation details, then tests the accuracy of DFT functionals. Sec.\,\ref{sec:results} presents our computational investigation of the multi-stage process; we address a) energy barriers for proton-hopping, b) vibrational properties and the pressure range of validity of the quasiharmonic approximation (QHA), c) order-disorder transition using \textit{mc}-QHA \cite{qinQhaPythonPackage2019}, its implications for the compressibility, and (001)-projected proton distribution d) neutron diffraction anomalies and disorder, and e) bond-length anomalies and tunneling. Sec.\,\ref{sec:conclusion} summarizes our results.

\section{Method}
\label{sec:method}

\subsection{DFT Details}

Our calculations were performed with the Quantum ESPRESSO (QE) code suite \cite{giannozziQUANTUMESPRESSOModular2009} using the Perdew-Burke-Ernzerhof (PBE) generalized gradient approximation (GGA) \cite{perdewGeneralizedGradientApproximation1996} to density functional theory (DFT). Evolutionary-optimized projector-augmented wave (EPAW) \cite{sarkarEPAW1CodeEvolutionary2018} datasets were used to describe valence-core electron interaction for all elements with electronic configurations $\mathrm{3s^23p^1}$, $\mathrm{2s^22p^4}$, and $\mathrm{1s^1}$ being used for Al, O, and H. The plane-wave cutoff energy was set to 80 Ry, and the Monkorst-Pack $k$-point grid for SCF calculations was set to $4 \times 4 \times 4$ for primitive unit cells and $4 \times 4 \times 2$ for $1 \times 1 \times 2$ supercells. At each pressure, dynamical matrices were obtained using density functional perturbation theory (DFPT) on a $2 \times 2 \times 2$ $q$-point grid and force constants were Fourier interpolated to produce phonon frequencies and vibrational density of states (VDoS) on an $8 \times 8 \times 8$ grid. The climbing-image nudged elastic band method (NEB) \cite{henkelmanClimbingImageNudged2000,henkelmanImprovedTangentEstimate2000} calculation is performed to determine the energy barrier between different configurations of HOC-\textgreek{δ}.

Standard DFT calculations do not easily describe systems with H-bonds. However, \textgreek{δ} consists of a 3D inter-connected Al-O-Al network formed by strong ionic bonds, which standard DFT functionals, both LDA (local density approximation) \cite{perdewSelfinteractionCorrectionDensityfunctional1981} and PBE \cite{perdewGeneralizedGradientApproximation1996}, describe well. Although a subtle distortion of the Al-O-Al network is caused by the H-bond \cite{umemotoPoststishoviteTransitionHydrous2016}, the general compressional mechanism of \textgreek{δ} is established by the Al-O bond compressibility and should be well described by both these functionals. However, the O—H···O configurations are more sensitive to the choice of functionals. For example, the H-bond (H···O) and ionic bond (O—H) lengths are expected to be overestimated and underestimated, respectively, as in ice VII and VIII \cite{umemotoOrderDisorderPhase2010, umemotoNatureVolumeIsotope2015}, more so with LDA than with PBE. Since this paper primarily addresses the H-bond, we choose to use the PBE-GGA functional.

The validity of the DFT functional can be tested by comparing pressure, i.e., equation of state (EoS), axial lengths, and interatomic distances vs.\ volume with experimental data. Test calculation results in Figs.\,S1 and S2 refer to the structures shown in Fig.\,\ref{fig:1}, which will be discussed below. For comparison, LDA results obtained with ultrasoft pseudopotentials \cite{garrityPseudopotentialsHighthroughputDFT2014} are also shown. However, HOC-12 and 11* configurations are unstable with LDA in the volume range of interest and are not shown. There are imaginary phonon frequencies in this volume range. As seen in Fig.\,S1, static LDA calculations underestimate and PBE overestimate pressure \cite{wentzcovitchThermodynamicPropertiesPhase2010}, with a difference of $\sim$5~GPa. Equilibrium axial lengths shown in Fig.\,S2 are well predicted by both functionals in static calculations, better by PBE. They are not affected much by H-bond configurations. LDA predicts $d$(AlO) better than PBE: the Al-O-Al network is more distorted with PBE. But more importantly, $d$(OH) is well-predicted by PBE (see Fig.\,S2(e)); LDA overestimates and underestimates $d$(OH) ionic- and H-bonds, respectively (see Fig.\,S2(f)). Therefore, we prefer the PBE functional to study the H-bond disorder and H-bond symmetrization process of \textgreek{δ}-AlOOH.

\subsection{Thermodynamics}

Multi-configuration QHA (\textit{mc}-QHA) calculations were performed with the open-source Python code \texttt{qha} \cite{qinQhaPythonPackage2019}. The QHA Helmholtz free energy for each configuration is given by
\begin{widetext}
\begin{equation}
F(V,T) = E_0(V) + \frac{1}{3N} \left\{
\sum_{\mathbf{q}s} \tfrac{1}{2}\hbar\omega_{\mathbf{q}s}(V) +
\sum_{\mathbf{q}s} k_\mathrm{B}T \ln\left[
    1 - \exp\left(-\frac{ \hbar\omega_{\mathbf{q}s}(V) }{ k_\mathrm{B}T }\right)
\right]\right\} .
\end{equation}
\end{widetext}
The multi-configuration QHA partition function $Z_\mathrm{QHA} (V,T)$ is then obtained by considering multiple inequivalent configurations with free energy $F_m (V,T)$ (Eq. (1)) and multiplicity $g_m$ \cite{umemotoOrderDisorderPhase2010, umemotoFirstprinciplesInvestigationHydrous2011}:
\begin{equation}
    Z_\mathrm{QHA} = \sum_m g_m \exp\left[ -\frac{F_m(V,T)}{k_\mathrm{B} T} \right] .
\end{equation}
The multi-configuration free energy $F_\mathrm{QHA} (V,T)$ is calculated from $Z_\mathrm{QHA}$ (V,T) as
\begin{equation}
    F_\mathrm{QHA} (V,T) = k_\mathrm{B} T \ln Z_\mathrm{QHA} (V,T) .
\end{equation}
The population of the $m$-th set of configurations, $n_m (V,T)$, is given by \cite{umemotoOrderDisorderPhase2010, umemotoFirstprinciplesInvestigationHydrous2011}:
\begin{equation}
    n_m(V,T) = \frac{Z_m}{Z_\mathrm{QHA}} = \frac{g_m\exp\left[-\frac{F_m(V,T)}{k_\mathrm{B}T}\right]}{\sum_m g_m\exp\left[-\frac{F_m(V,T)}{k_\mathrm{B}T}\right]} .
\end{equation}
Ensemble average of property ($A(V,T)$), e.g., diffraction intensity $I_{021}$, bond $d$(OH), are computed as an average of $A_m$ weighted by the population $n_m$, i.e.,
\begin{equation}
    A(V,T) = \sum_m n_m(V,T) \cdot A_m(V,T) .
\end{equation}
All quantities (e.g., $n_m$, $A$) can then be converted from a function of $(V,T)$ to a function of $(P,T)$ using $V(P,T)$, i.e.,
\begin{equation}
    n_m(P,T) = \left.n_m(V,T)\right|_{V = V(P,T)} ,
\end{equation}
\begin{equation}
    A(P,T) = \left.A(V,T)\right|_{V = V(P,T)} ,
\end{equation}
where $P(V,T)= -\left(\partial F_\mathrm{QHA} / \partial V\right)_T$.

This method has been successfully applied to obtain the thermodynamic phase boundary between H-bond disordered ice-VII and ice-VIII \cite{umemotoOrderDisorderPhase2010}, to study hydrous defects in forsterite \cite{qinInitioStudyWater2018, umemotoFirstprinciplesInvestigationHydrous2011}, and to study an order-disorder transition in a low-pressure analog \cite{umemotoInitioPredictionOrderdisorder2021} of an ultra-high-pressure planet forming silicate \cite{umemotoPhaseTransitionsMgSiO32017}.

\subsection{Structure}

\begin{figure}[t]
    \centering
    \includegraphics[width=.48\textwidth]{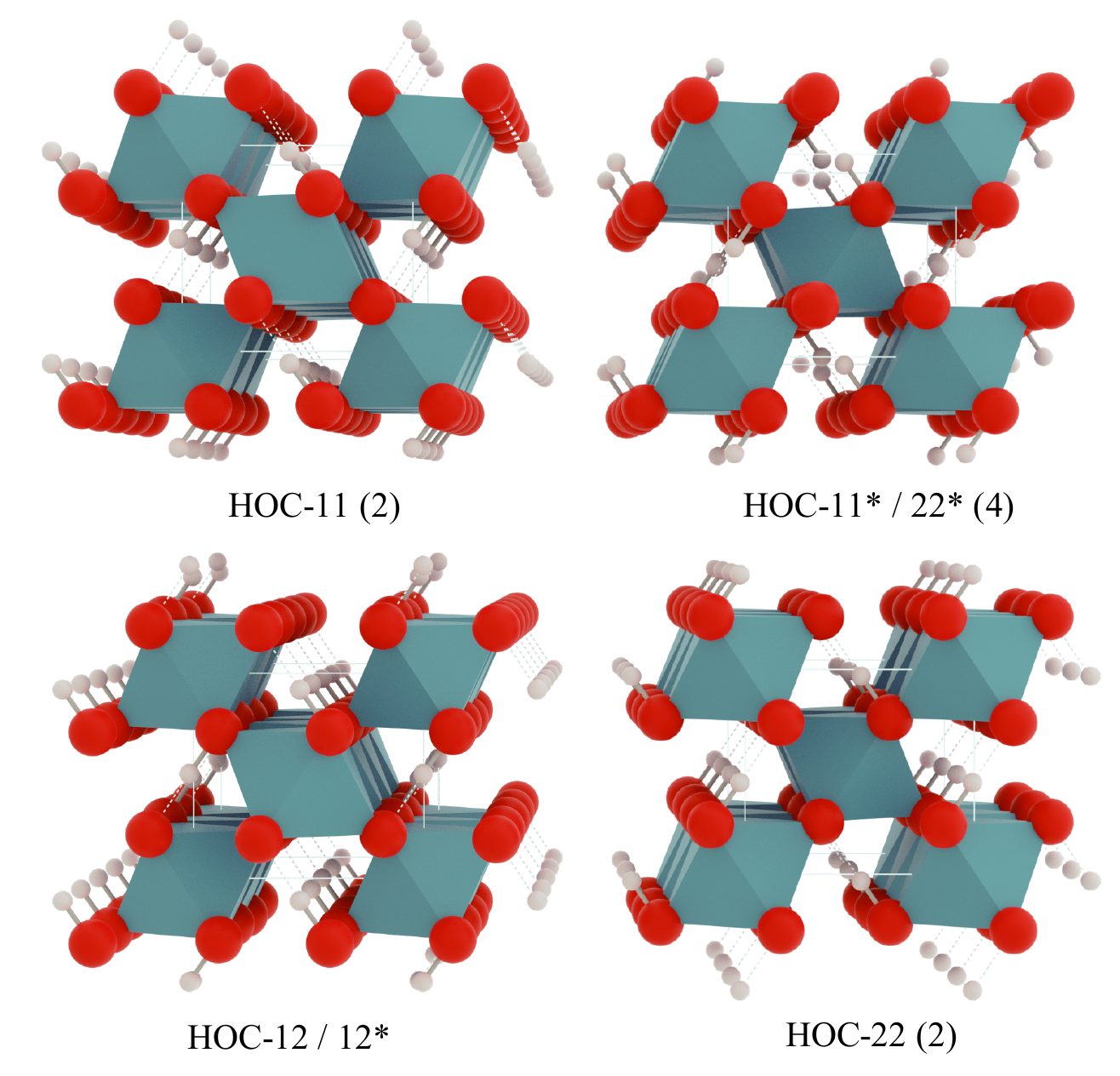}
    \caption{Views along the $z$-direction of structure models of \textgreek{δ}-AlOOH $1 \times 1 \times 2$ supercells investigated in this study. The numbers in parentheses denote their multiplicities. These are the same ones as those proposed by Ref.\,\cite{tsuchiyaVibrationalPropertiesDAlOOH2008}, although 11* and 22*, 12 and 12* in Ref.\,\cite{tsuchiyaVibrationalPropertiesDAlOOH2008} are symmetrically equivalent. Red and white spheres denote O and H ions, respectively. Teal octahedra represent Al—O coordination octahedra. Short solid and long white lines indicate O—H ionic and O···H H-bonds, respectively.}
    \label{fig:1}
\end{figure}

\textgreek{δ}-AlOOH’s primitive cell contains two formula units (f.u.). Except for results in Sec.\,\ref{sec:3A}, this study will model H-bond disorder with $1 \times 1 \times 2$ supercells containing four \textgreek{δ}-AlOOH f.u.. Fig.\,\ref{fig:1} shows four symmetrically distinct $1 \times 1 \times 2$ supercell structures included in this study. These are the same supercells proposed by Tsuchiya et al. \cite{tsuchiyaVibrationalPropertiesDAlOOH2008}, except that duplications (i.e., 11*/22* and 12/12*) are merged. The $1 \times 1 \times 2$ supercells can have 16 ($2^4$) configurations in total, which reduce to the four symmetrically distinct configurations shown in Fig.\,\ref{fig:1} with multiplicities 2, 4, 8, and 2. Any other configuration in larger supercells can be described as a combination of these four sets of configurations in variable amounts. The interaction energy between these configurations is not included in \textit{mc}-QHA calculations.

\subsection{Pressure notation}

At the same volume, PBE/\textit{mc}-QHA calculations generally overestimate pressure by $\sim$5~GPa compared to measurements. To avoid confusion, we use the \unit{GPa}$^\text{EXP/PBE/QHA}$ unit to refer to pressure from 300~K measurements, static PBE calculations, and 300~K QHA or \textit{mc}-QHA calculations in subsequent discussions.

\section{Results and Discussion}
\label{sec:results}

\subsection{Proton jump energy barrier vs.\ pressure}
\label{sec:3A}

\begin{figure}[t]
    \centering
    \includegraphics[width=0.48\textwidth]{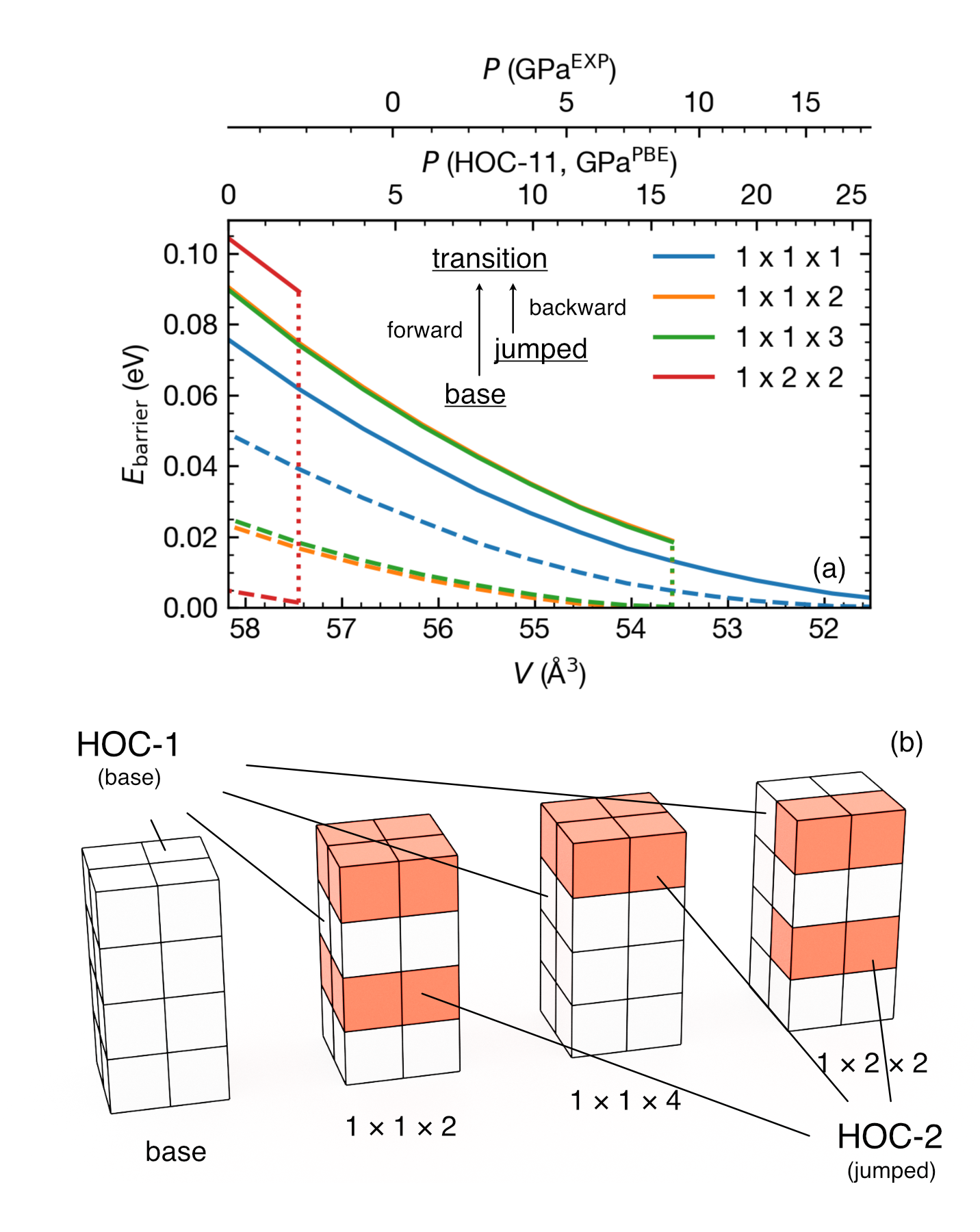}
    \caption{(a) Volume dependence of NEB activation energies, or energy barrier heights for a single (but periodically repeated) proton jump from HOC-1 to HOC-2. Dashed and solid lines indicate activation energies for forwarding and backward jumps; colors denote the supercell size in the NEB calculation. Green lines generally overlap with yellow ones. The bottom and top axes show the unit-cell volume (2 f.u.) and corresponding experimental pressure and static PBE pressure of the base model (HOC-1) for reference. (b) Base and jumped structures used in the NEB calculations for different supercell sizes are shown in a $2 \times 2 \times 4$ supercell. The supercells are constructed by stacking HOC-1 (base) and HOC-2 (jumped) unit cells indicated by white and orange cubes. HOC-1 and HOC-2 configurations differ by a single change in proton configuration, i.e., a single proton jump per primitive cell. }
    \label{fig:2}
\end{figure}

Before investigating proton disordering in \textgreek{δ}, we use NEB to calculate the activation energy required for a single proton jump (or energy barrier) and its pressure dependence. There are two protons in each HOC-1 or HOC-2 unit cell configuration of \textgreek{δ} (HOC-11 and HOC-22 supercells in Fig.\,\ref{fig:1}). A single periodic proton jump in the unit cell from the H1 to the H2 site results in a “forward” structural change (from HOC-1 to HOC-2); a jump from H2 to H1 results in a “backward” change. The activation energies for the forward and backward periodic jumps in a single unit cell are shown in Fig.\,\ref{fig:2}(a) as solid and dashed blue curves, respectively. At 0~\unit{\GPaPBE}, the activation energy is $\sim$0.075~eV ($\sim$870~K). Under compression, the energy barrier decreases significantly to roughly 0.025~eV ($\sim$300~K) at $\sim$11~\unit{\GPaPBE} and almost vanishes at $\sim$20~\unit{\GPaPBE}. The barrier height reduction justifies the evolution of proton distribution in a classical Langevin molecular dynamic (MD) analysis that shows a change in OH radial pair-correlation distribution (Fig.\,\ref{fig:4} in Ref.\,\cite{bronsteinThermalNuclearQuantum2017}) from discrete at 0~\unit{\GPaPBE} to interconnected at $\sim$10~\unit{\GPaPBE} \cite{bronsteinThermalNuclearQuantum2017}.

A path integral MD analysis of proton motion shows that protons in H-bonds would be able to tunnel through the barrier wall roughly when the barrier height between neighboring proton sites become comparable to $k_\mathrm{B} T$, the thermal energy \cite{benoitShapesProtonsHydrogen2005}. Our calculations show that at 0~\unit{\GPaPBE}, the energy barrier for a forward jump is larger than $\sim$0.026~eV ($\sim$300~K). But at higher pressure, the barrier decreases to less than $k_\mathrm{B} T$ at 300~K, showing that \textgreek{δ} could enter a strongly anharmonic and tunneling regime before or around 11~\unit{\GPaPBE}. From a statistical perspective, $p_\mathrm{jump} / p_\mathrm{base} = \exp(-\Delta E / k_\mathrm{B}T)$ where $\Delta E = E_\mathrm{jump} - E_\mathrm{base})$. At 300~K and 0~\unit{\GPaPBE}, $\Delta E \sim 0.08$ eV and $p_\mathrm{jump} / p_\mathrm{base} \sim 1/25$; at 11~\unit{\GPaPBE}, $p_\mathrm{jump} / p_\mathrm{base} \sim 1/2.7$ only. At higher pressures and temperature ($P,T$) conditions of subducted slabs \cite{litasovPhaseRelationsHydrous2005}, it is entirely possible for the structure to disorder (as described below).
To simulate a solitary proton jump in a crystal more accurately and understand size effects, we model a single proton jump in various supercells. They are $1 \times 1 \times 2$, $1 \times 1 \times 4$, and $1 \times 2 \times 2$ multiples of a unit cell. Fig.\,\ref{fig:2}(b) represents the “base” and “jumped” configurations. For one jump on a $1 \times 1 \times 2$ supercell, the initial and final structures correspond to HOC-11 and HOC-12 in Fig.\,\ref{fig:1}. The energy barrier increases for a forward jump and decreases for a backward jump. At $\sim$17~\unit{\GPaPBE}, the backward energy barrier decays to zero. Other $1 \times 1 \times 2$ supercell final configurations are possible but correspond to multiple jumps. For example, the transition from HOC-11 to HOC-11* (-11, -11*, etc., subsequently) structures in Fig.\,\ref{fig:1} can be achieved by jumping twice, -11→-12→-11*. Further increase in supercell size along the $c$ direction does not result in substantial differences in energy barrier (e.g., in $1 \times 1 \times 2$ or $1 \times 1 \times 3$, etc). 

Proton jumps on other types of supercells offer surprising results. We are unable to find a stable static structure with a single jump in (001), the $a, b$-plane, e.g., in $1 \times 2 \times 1$ or $2 \times 1 \times 1$ supercells. The $1 \times 2 \times 2$ supercell can produce a single jump final structure at $\sim$0~\unit{\GPaPBE}, but only in a narrow pressure range (see Fig.\,\ref{fig:2}(a)). Even in this narrow range, the energy barrier for backward jump decreases substantially compared to the other $1 \times 1 \times N$ jumped supercell configurations. This result means an isolated proton jump in (001) is unlikely to result in a long-lived state. This result confirms that a single proton jump in an ordered structure is not likely, and proton jumps must be coordinated in the $a, b$-plane. Even in H-disordered \textgreek{δ}, periodicities along the a and b axes are likely to be maintained in some length scale. For this reason, we must consider a more orderly type of disorder. Dynamically, the change in H-bond configuration (disorder) in the $a, b$-plane might occur in sequential jumps as in-plane soliton propagation, which is a consequence of the “one proton per octahedron” limit imposed by the chemistry.

Because the size effect is sufficiently mitigated with supercells with two units along [001], and periodicity should remain mostly intact in the (001) plane, at least within a finite length scale, our results suggest that $1 \times 1 \times 2$ supercell calculations offer ample base to address structural disorder in \textgreek{δ}. Moreover, since Ref.\,\cite{tsuchiyaVibrationalPropertiesDAlOOH2008} has confirmed that the Raman spectrum in the OH-stretching region is generally reproduced with these units, the ensemble of all possible $1 \times 1 \times 2$ supercells might suffice to describe all possible short-range O—H···O configurations. The six structures given in Ref.\,\cite{tsuchiyaVibrationalPropertiesDAlOOH2008} can be reduced to four independent ones shown in Fig.\,\ref{fig:1}, HOC-11, HOC-11*/22*, HOC-12, and HOC-22 (11*/22* and 12/12* reported in Ref.\,\cite{tsuchiyaVibrationalPropertiesDAlOOH2008} are symmetrically related). The remaining sections of this study use these supercells to investigate the compressional behavior of H-bonds in \textgreek{δ}.

\subsection{Pressure dependence of VDoS, proton stability, and tunneling signature}
\label{sec:3B}

After confirming that these supercell models describe well the predominant short-range proton configurations, it is essential to clarify the pressure dependence of their related phonon frequencies and modes. The high energy barriers discussed in the previous section suggest minor anharmonic effects on proton motion at low $P,T$s. However, these modes should become anharmonic with increasing $P,T$s. Therefore, the highest-pressure results reported in this section have more phenomenological and qualitative significance than predictive power. Nevertheless, they offer a detailed view of trends.

\begin{figure*}
    \centering
    \includegraphics[width=\textwidth]{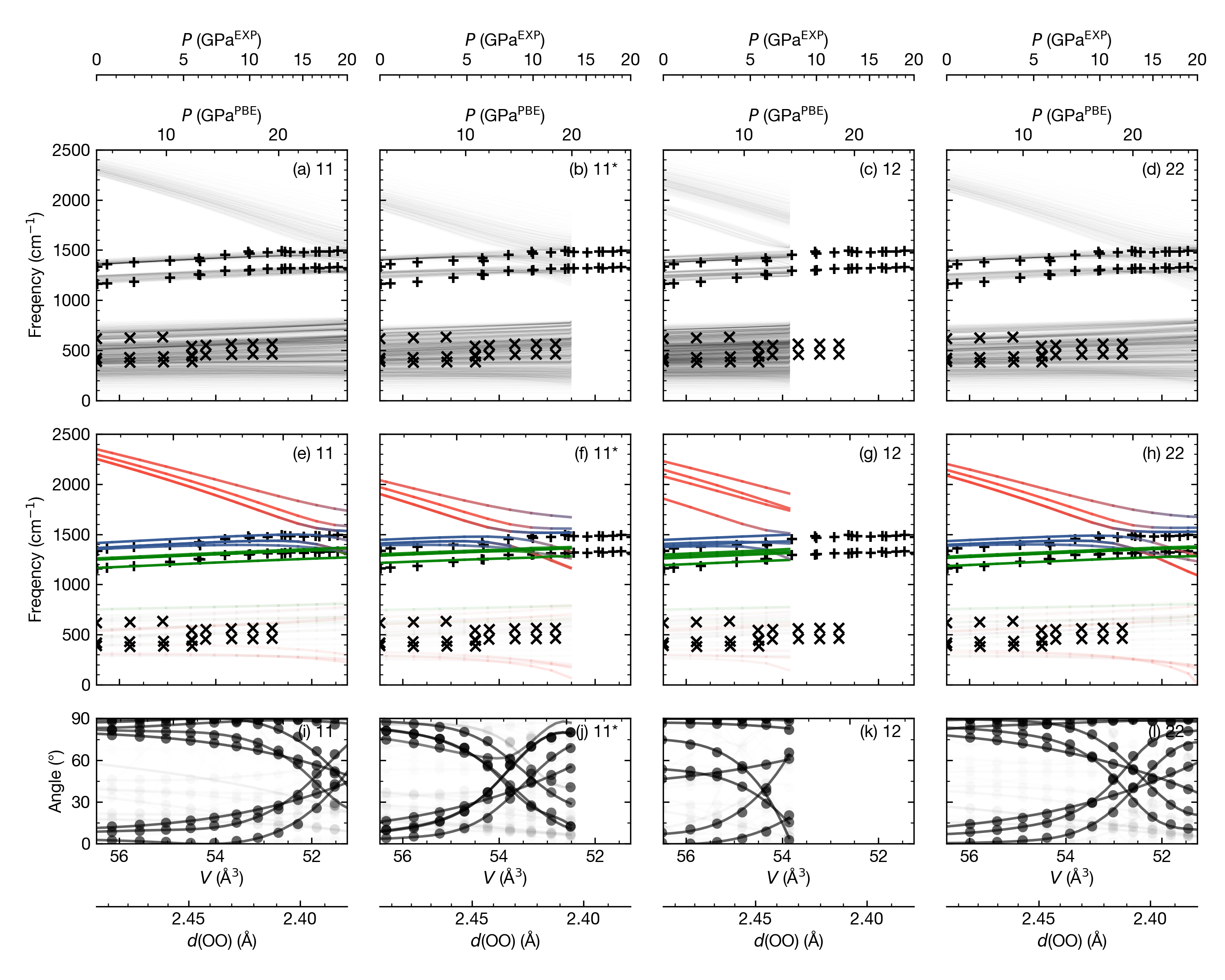}
    \caption{Pressure evolution of phonon properties. (a–d) H-projected vibrational density of states (H-VDoS) vs.\ unit-cell volume. The transparency of color indicates VDoS amplitude; (e–h) \textgreek{Γ} point phonon frequencies of OH stretching (red), in-plane bending (blue), and out-of-plane bending (green) modes vs.\ unit cell volume. The transparency of color indicates projected strength; (i–l) angle between H-displacement vector and OH bond direction vs.\ volume. Symbols are from PBE results, and lines are spline interpolations to help identify mode-continuity with volume. The main bottom axis indicates the unit cell volume (2~f.u.); Extra axes at the top show the experimental and PBE static pressure at the investigated volumes. The additional axis at the bottom, $d$(OO) from static PBE calculations, is given to facilitate discussions. Scattered “+” and “×” symbols in panels (a–h) denote the frequency of IR and Raman peaks extracted from Refs.\,\cite{kagiInfraredAbsorptionSpectra2010} and \cite{mashinoSoundVelocitiesDAlOOH2016}.}
    \label{fig:3}
\end{figure*}

Fig.\,\ref{fig:3} shows the evolution of various phonon properties under pressure. Because phonon frequency behavior is a function of volume or bond-length (see, e.g., \cite{umemotoNatureVolumeIsotope2015}), and our system of interest, O—H···O, is relatively localized and could be studied vs.\ $d$(OO) bond lengths \cite{benoitShapesProtonsHydrogen2005}, it is more informative to display frequencies vs.\ volume henceforth. As will be shown in Sec.\,\ref{sec:3E}, $d$(OO) is nearly a linear function of volume and is well-predicted. Therefore, we add a scale showing the corresponding PBE $d$(OO) bond-lengths. We also supply different scales at the top, indicating PBE and experimental pressures at corresponding volumes. Phonon frequency continuity between adjacent volume points is established using eigenvectors similarity described in Ref.\,\cite{luoCijPythonCode2021a}.

Figs.\,\ref{fig:3}(a–d) show the pressure evolution of the vibrational density of states (VDoS) obtained by first interpolating phonon frequency vs.\ volume, then calculating the VDoS at each volume on a dense volume grid for each of the four structures in Fig.\,\ref{fig:1}. Across the spectrum, the phonon bands cover three broad frequency domains: the 500–800~\unit{cm^{-1}} bands consist of Al and O (heavier-ions) related modes, the 1000–1500~\unit{cm^{-1}} bands are OH bending modes, and the 1500–2500~\unit{cm^{-1}} ones are OH stretching modes. The pressure dependence of the 500–800~\unit{cm^{-1}} modes is similar to those of stable anhydrous phases and overlaps with Raman mode frequencies \cite{mashinoSoundVelocitiesDAlOOH2016}. The high-frequency modes, both the 1000–1500~\unit{cm^{-1}} and the 1500–2500~\unit{cm^{-1}} bands, are more strongly pressure-dependent. The stretching and bending modes’ negative and positive pressure coefficients are similar to those reported in previous theoretical studies of \textgreek{δ} \cite{tsuchiyaVibrationalPropertiesDAlOOH2008, bronsteinThermalNuclearQuantum2017} and other H-abundant phases \cite{nakamotoStretchingFrequenciesFunction1955, umemotoTheoreticalStudyIsostructural2005}. The current static calculation slightly underestimates $d$(OO) and the stretching-mode frequencies compared to other works \cite{tsuchiyaVibrationalPropertiesDAlOOH2008, bronsteinThermalNuclearQuantum2017}. All phonon frequencies exhibit a near-linear dependence on volume until the OH-stretching and -bending mode frequencies become similar and these modes start interacting.

Comparing the VDoSs of the four supercell models reveals a meaningful relationship between the OH stretching band frequencies and proton arrangements along the interstitial channels. There are two stretching mode bands, one beginning above and one below 2000~\unit{cm^{-1}} at 0~\unit{\GPaPBE}. The former exists in the HOC-11, -22, and -12 configurations and the latter exists in HOC-11* and -12 (see Fig.\,\ref{fig:3}(a–d)). There are two types of proton arrangements: in HOC-11 and -22, protons are aligned along the interstitial channels parallel to the $c$ axis, i.e., their positions are related by a translation $c$. In HOC-11*, protons are half-aligned, i.e., their positions are related by a $2c$ translation. This proton arrangement is more “disordered”. The HOC-12 configuration has both aligned and half-aligned types of protons. Aligned protons have higher OH-stretching frequencies than half-aligned ones. This interpretation sheds light on the origin of results obtained with a 300~K quantum thermal bath (QTB) MD simulation performed on $2 \times 2 \times 4$ supercells \cite{bronsteinThermalNuclearQuantum2017}. Two series of OH-stretching modes in the 0–10~GPa pressure range were reported indicating the coexistence of aligned and half-aligned arrangements. The disappearance of the higher-frequency band at 10~GPa can be attributed to the disappearance of the aligned proton configurations, indicative of growing disorder in the structure. Mostly HOC-11*-like arrangements are expected to be present beyond 10~GPa in those simulations. This configuration change will be addressed with \textit{mc}-QHA calculations in Sec.\,\ref{sec:3C}.

Figs.\,\ref{fig:3}(e–h) show the evolution of OH zone center (\textgreek{Γ}-point) mode frequencies under compression. Red, green, and blue denote the OH-stretching, OH-bending in the $a, b$-plane (in-plane bending), and OH-bending along the $c$-direction (out-of-plane bending). This classification of proton modes is based on local mutually perpendicular axes defined by (a) the in-plane ionic O—H bond direction, (b) the perpendicular direction in (001), the $a, b$-plane, and (c) a third out-of-plane direction perpendicular to both, i.e., along [001]. The intensity of color represents the magnitude of eigenmode projections onto these OH displacement directions. Translucent curves mean the proton displacement in those modes correlates strongly with Al or O ionic motions. 

The four proton configurations are stable in different pressure ranges in static \textit{ab initio} calculations. When a proton configuration becomes unstable, it transforms into another one. Imaginary frequencies in Figs.\,\ref{fig:3}(e–h) identify proton vibrational instabilities at the \textgreek{Γ} point. HOC-12, -11*, and -22 develop signs of vibrational instability at $\sim$12, 22, and 26~\unit{\GPaPBE}, or $\sim$8, 12, and 20~\unit{\GPaEXP}, while HOC-11 survives beyond 26~\unit{\GPaPBE}. Figs.\,\ref{fig:3}(e–h) show that unstable modes have a slightly reddish color, indicating the onset of OH tunneling behavior (stretching mode instability), similar to earlier reports \cite{bronsteinThermalNuclearQuantum2017}.

As for the high-frequency modes, we observe a “mode mixing” and “frequency crossing” between stretching and in-plane OH-bending modes preceding the vibrational instabilities in all four supercells shown in Figs.\,\ref{fig:3}(e–h). As stretching mode frequencies approach the in-plane bending frequencies, they repel each other, these modes mix, and the stretching and bending branches start to switch order. This process is continuous, and the eigenmodes neither stretch nor bend throughout a few GPa while these frequencies overlap. Similar mixing behavior was previously reported in these modes’ harmonic analysis \cite{tsuchiyaVibrationalPropertiesDAlOOH2008, bronsteinThermalNuclearQuantum2017}. An OH-stretching bandwidth increase (see Figs.\,\ref{fig:3} (a–d)) accompanies this “mode mixing”.

This mode “mixing” process is captured in the volume dependence of the angle between the O—H displacement vector and the O—H bond direction shown in Figs.\,\ref{fig:3}(i–l). Fig.\,S3 shows in more detail the pressure evolution of these eigenmodes in HOC-11*. Over the entire process, the displacement directions of both modes undergo a continuous 90° rotation. We do not observe any strong distortion in the O-octahedra throughout this process. This switch in the mode frequency order results from the contraction of interstitial site volume or H-bond length under pressure. A similar phenomenon was reported in MD simulations of disordered ice-VII at $d$(OO) around 2.4~Å \cite{benoitShapesProtonsHydrogen2005}, and it could be a general phenomenon in O—H···O bearing systems. In ice-VII \cite{benoitShapesProtonsHydrogen2005}, this mode crossing behavior was described as a pre-tunneling effect, suggesting that \textgreek{δ} with specific H-bond configurations will enter a similar regime at corresponding pressure (e.g., 11~\unit{\GPaPBE} or 8~\unit{\GPaEXP} for HOC-12, 16~\unit{\GPaPBE} or 10.5~\unit{\GPaEXP} for HOC-11*). We will address this issue further in the upcoming sections under the context of proton configuration population evolution from the \textit{mc}-QHA analysis.

With a better understanding of the pressure dependence of eigenmodes and their frequencies, we can reinterpret the pressure dependence of IR frequencies \cite{kagiInfraredAbsorptionSpectra2010} and tentatively connect them with disordering, tunneling, and symmetrization in subsequent sections. When the measured IR OH-bending mode frequencies \cite{kagiInfraredAbsorptionSpectra2010} shown in Fig.\,\ref{fig:3} are more clearly displayed in Fig.\,S4, various regions with different frequency vs.\ volume dependencies are better identified. Slope changes indicate changes in the proton arrangement or vibrational state in \textgreek{δ}. For in-plane OH-bending frequencies, the first slope change happens at $\sim$7.5–8~\unit{\GPaEXP}. There is an abrupt compressibility increase \cite{sano-furukawaChangeCompressibilityDAlOOH2009}, and the 021 neutron diffraction peak disappears \cite{sano-furukawaDirectObservationSymmetrization2018, kuribayashiObservationPressureinducedPhase2014} at this same pressure. It also happens that the HOC-12 configuration becomes vibrationally unstable at this pressure (see Figs.\,\ref{fig:3}(c,g,k)). The disappearance of the 2800~\unit{cm^{-1}} branch at 10~GPa in MD simulations \cite{bronsteinThermalNuclearQuantum2017} should also correspond to this change because the HOC-12 configuration essentially disappears after this point. The slope change in the out-of-plane OH-bending frequency happens later, at $\sim$11~\unit{\GPaEXP} (see Fig.\,S4(a)). A kink is observed immediately after at 11.5~\unit{\GPaEXP}, marking the beginning of the 11.5–18~\unit{\GPaEXP} range in which $d$(OH) distances behave anomalously \cite{sano-furukawaDirectObservationSymmetrization2018} (see Fig.\,S4(b)). We correlate this change with the anharmonicity in the HOC-11* type structure which starts at $\sim$18~\unit{\GPaPBE} ($\sim$11.5~\unit{\GPaEXP}). Finally, at 18~\unit{\GPaEXP}, we observe slope changes in both in-plane and out-of-plane bending frequencies (see Fig.\,S4). This pressure corresponds to the full H-bond symmetrization pressure \cite{sano-furukawaDirectObservationSymmetrization2018}. These connections will be explored in greater detail in subsequent sections and give fresh insights into the nature of this complex multi-stage phase change in \textgreek{δ} complicated by proton tunneling.

\subsection{Population analysis, thermodynamic properties from \textit{mc}-QHA, 300~K EoS, implications for compressibility, and proton distribution}
\label{sec:3C}

The small ($< 0.1$ eV/f.u.) energy differences between the four supercell H-bond configurations shown in Fig.\,\ref{fig:1} suggest the disordered system is indeed possible \cite{tsuchiyaVibrationalPropertiesDAlOOH2008}. We now investigate the thermodynamic properties of HOC-\textgreek{δ} using the \textit{mc}-QHA method \cite{qinQhaPythonPackage2019} and the volume dependence of VDoSs of the four supercells. As previously discussed in Sec.\,\ref{sec:3A}, the four structures shown in Fig.\,\ref{fig:1} represent well possible short-range proton configurations. Their coexistence in different proportions describes structural disorder in our model.

Because QHA calculations require stable phonons, four separate \textit{mc}-QHA calculations were performed in different pressure ranges (0–12, 12–20, 20–26, and 26–30~\unit{\GPaQHA}). Different proton configurations are stable in these pressure ranges. Since phonons are calculated assuming harmonic potentials, anharmonic effects and proton tunneling are not accounted for in this treatment. The current \textit{mc}-QHA approach can shed light on disordering effects and H-bond symmetrization but is not predictive.

\begin{figure}
    \centering
    \includegraphics[width=.45\textwidth]{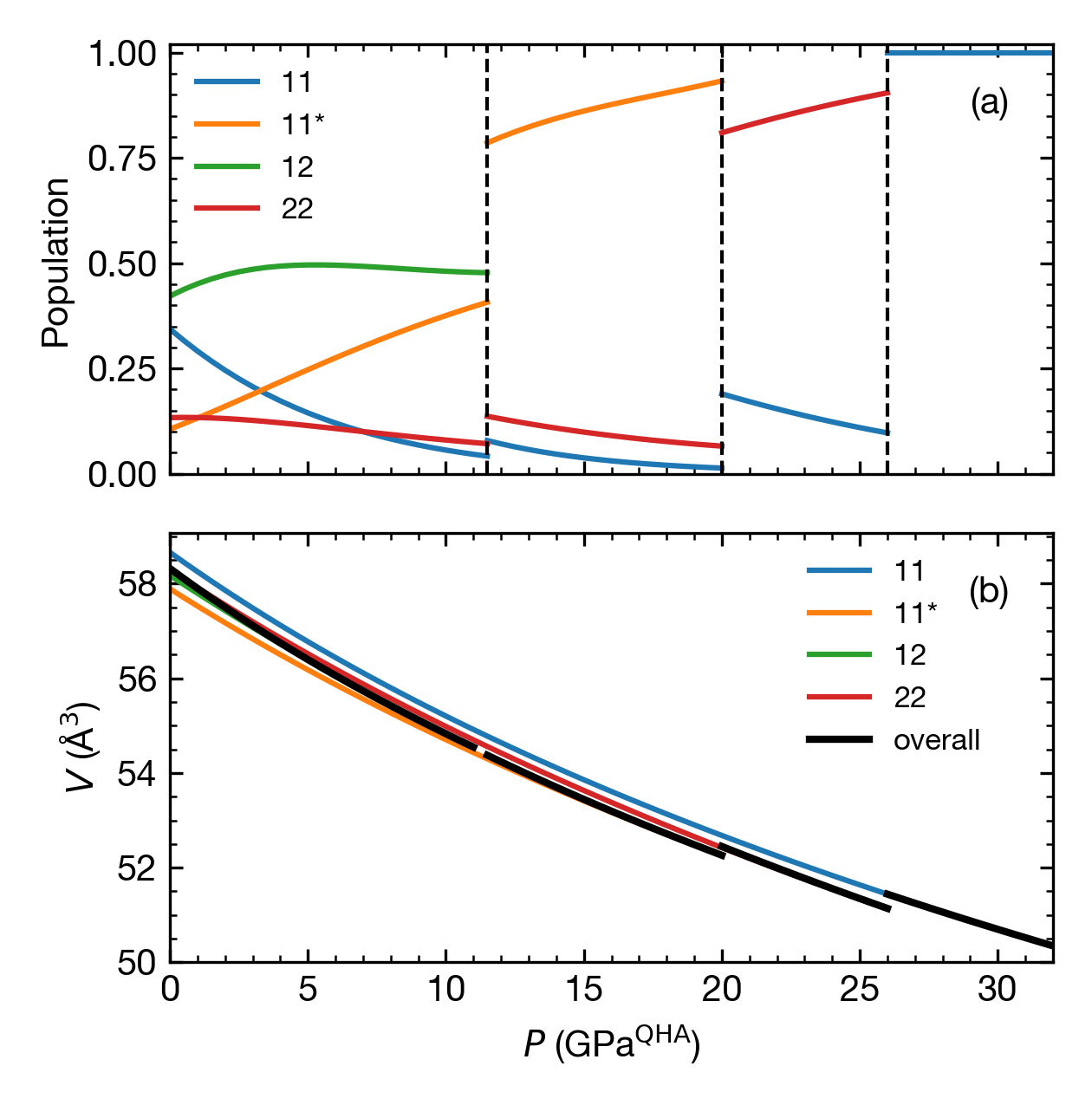}\\
    \includegraphics[width=.45\textwidth]{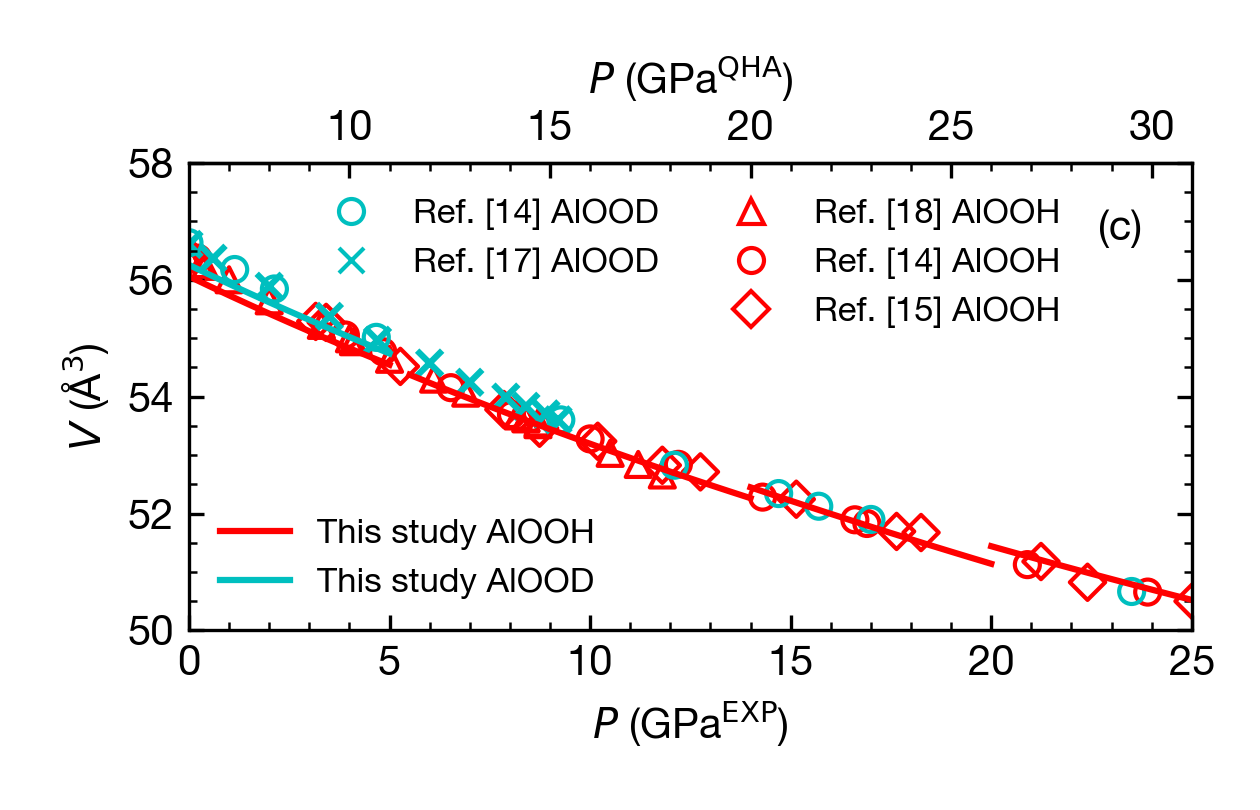}
    \caption{(a) Pressure evolution of HOC-11, -11*, -12, and -22 populations from \textit{mc}-QHA at 300~K.  (b) Comparison between HOC-11, -11*, -12, and -22 EoSs and the multi-configuration overall volumes (2~f.u.).  (c) Comparison between the overall \textgreek{δ}-AlOOH (red) and \textgreek{δ}-AlOOD (blue) volumes at 300~K. “Overall” volumes are from \textit{mc}-QHA calculations. Circles, diamonds, crosses, and triangles denote experimental data from \cite{sano-furukawaChangeCompressibilityDAlOOH2009}, \cite{simonovaStructuralStudyDAlOOH2020}, \cite{sano-furukawaNeutronDiffractionStudy2008}, and \cite{kuribayashiObservationPressureinducedPhase2014}, respectively. Pressure from \textit{mc}-QHA (top axis) in (c) is shifted by –6~GPa to emphasize similarities between experimental and \textit{mc}-QHA compression curves.}
    \label{fig:4}
\end{figure}

In this section, thermodynamic properties are discussed at 300~K and QHA pressure. However, our 300~K QHA calculation systematically overestimates pressure by $\sim$4–6~GPa compared to experiments at the same volume due to the use of the PBE exchange-correlation functional and other factors, e.g., anharmonicity (see Fig.\,\ref{fig:4}(c) and Fig.\,S1). Therefore, a shift in pressure is needed when comparing our proposed boundary pressures with experimental pressures (e.g., $-6$ GPa shift applied to Fig.\,\ref{fig:4}(c)). This pressure shift is also required when comparing measured and calculated $d$(OO) bond lengths.

Fig.\,\ref{fig:4}(a) shows the evolution of different HOC configuration populations vs.\ pressure at 300~K. In the 0–12~\unit{\GPaQHA} range, the HOC-12 configuration is the most abundant, the HOC-11’s population decreases, HOC-22’s remains low, and HOC-11*’s increases with pressure. The vibrational instability in the HOC-12 structure at $\sim$12~\unit{\GPaQHA} leads to the disappearance of this atomic arrangement beyond this pressure. In the 12–20~\unit{\GPaQHA} range, HOC-11* becomes the most abundant configuration, and the HOC-11’s and HOC-22’s populations remain low. After HOC-11* becomes unstable at $\sim$20~\unit{\GPaQHA}, HOC-22 and -11 configurations are still stable in the 18–26~\unit{\GPaQHA} range. In the 26–30~\unit{\GPaQHA} range, near H-bond symmetrization, only HOC-11 seems to survive, but HOC-22 and HOC-11 are the same when the H-bond is symmetric.

A closer look at the free energy of these structures (Fig.\,S5) sheds light on the driving force behind this structural evolution. We showed in Sec.\,\ref{sec:3B} that H-bond arrangements affect stretching mode frequencies. Fig.\,S5 shows how these differences in vibrational properties contribute to the free energy, in particular, the zero-point motion energy ($E_\mathrm{zp}$). Because H-bond arrangements half-aligned along the [001] have lower stretching mode frequencies, half-aligned proton configurations have lower $E_\mathrm{zp}$. Although HOC-11 has the lowest static energy/enthalpy, the vibrational energy contribution becomes more important as differences in static energy decrease under compression. As shown in Fig.\,S5, from $\sim$5~\unit{\GPaPBE} onwards (or $\sim$1~\unit{\GPaEXP}), HOC-11* and -11 become the most and least stable configurations, respectively, owing to $E_\mathrm{zp}$. With mixed proton configurations, HOC-12 is not at the extremes of this energy spectrum. Still, its greater multiplicity makes it the most populous configuration in the 0–12~\unit{\GPaEXP} range until it becomes unstable at 12~\unit{\GPaEXP}. After 12~\unit{\GPaEXP}, the HOC-11* configuration is the most abundant due to its energy and multiplicity. Essentially, the decreasing static free-energy difference in the disordered system allows the half-aligned proton configurations to be more populous at higher pressures after the HOC-12 configuration becomes unstable. In summary, disregarding anharmonicity or dynamic disorder, the current \textit{mc}-QHA calculation suggests the transitions in HOC-\textgreek{δ} occur in the following order: in 0–8~\unit{\GPaEXP} (0–12~\unit{\GPaQHA}) \textgreek{δ} undergoes a continuous change in H-bond configuration with HOC-11 transforming into -11*; at 8~\unit{\GPaEXP} (12~\unit{\GPaQHA}), HOC-12 transforms to -11*; at 14~\unit{\GPaEXP} (20~\unit{\GPaQHA}), HOC-11* transforms to -22; at 20~\unit{\GPaEXP} (26~\unit{\GPaQHA}), HOC-22 transforms to -11. Because of multiplicity and zero-point energy, the 0~\unit{\GPaEXP} measured Raman spectrum \cite{ohtaniStabilityFieldNew2001} resembles the calculated HOC-12 spectrum more closely than the -11 \cite{tsuchiyaVibrationalPropertiesDAlOOH2008}, the most stable configuration at this pressure. This underlines the necessity of including configuration multiplicity and vibrational contributions in \textit{ab initio} calculations of H-bond disordered systems.

The retirement of the aligned protons in the HOC-12 configuration at 8~\unit{\GPaEXP}, i.e., the instability in HOC-12 and the dominance of the -11* configuration afterward, is consistent with the disappearance of high-frequency stretching bands at $\sim$10~GPa in 300~K QTB MD simulations \cite{bronsteinThermalNuclearQuantum2017}. At least within a limited pressure range, i.e., up to 12~\unit{\GPaEXP} (18~\unit{\GPaQHA}), the predominance of the half-aligned proton configuration in HOC-11* is confirmed by both \textit{mc}-QHA and QTB MD simulations at 300~K \cite{bronsteinThermalNuclearQuantum2017}. After 12~\unit{\GPaEXP} (18~\unit{\GPaQHA}), \textit{mc}-QHA calculations and QTB simulations at 300~K \cite{bronsteinThermalNuclearQuantum2017} no longer offer consistent results, most likely because \textit{mc}-QHA calculations cannot properly capture anharmonic effects. In the QTB MD simulations \cite{bronsteinThermalNuclearQuantum2017}, no discontinuity is observed in the phonon stretching band, which is expected if the half-aligned proton configuration in HOC-11* vanishes and the proton-aligned configurations in HOC-11 and -22 become dominant as predicted by \textit{mc}-QHA. Despite the softening of branches approaching imaginary frequencies (Fig.\,\ref{fig:3}(f)), we can still optimize the HOC-11* structure or a distorted version in static calculations. This signals a small energy barrier (Fig.\,\ref{fig:2}(a)) that, in reality, might fail to trap protons locally due to tunneling or anharmonic fluctuations. 

This population evolution vs.\ pressure significantly affects the 300~K EoS and compressibility of the multi-configuration system, as seen in Fig.\,\ref{fig:4}(b). This figure also shows the 300~K compression curves for each of the four configurations for comparison. Their volumes differ by less than 0.4~\unit{\cubic\angstrom} at any pressure and are in the following order: $V_{11} > V_{22} > V_{12} > V_{11^*}$. Intuitively, the half-aligned H-bond configurations keep protons farther apart, lower their stretching mode frequencies, and reduce the volume. 

Piecewise, the overall compression curve predicted by \textit{mc}-QHA resembles more closely that of the predominant configuration in different pressure ranges (see Fig.\,\ref{fig:4}(c)). In the 0–12~\unit{\GPaQHA} (0–9~\unit{\GPaEXP}) range, the decrease in population of aligned configurations (HOC-11 and -22) and the increase of half-aligned ones (HOC-12 and -11*) results in an increase in compressibility w.r.t.\ that of any of the configurations alone (see Fig.\,\ref{fig:4}(b)). The same happens with the extinction of HOC-12 at 12~\unit{\GPaQHA}. This result correlates well with the measured high compressibility regime (0–9~\unit{\GPaEXP}) before the anomalous stiffening \cite{sano-furukawaChangeCompressibilityDAlOOH2009}.

In a limited pressure range of 12–20~\unit{\GPaQHA} or 9–14~\unit{\GPaEXP}, HOC-11* alone determines the shape of the compression curve. In this pressure range, the compressibility decreases compared to that in the 0–12~\unit{\GPaQHA} range (see Fig.\,\ref{fig:4}(b)). This behavior correlates with the observed behavior in the 9–12~\unit{\GPaEXP} pressure range \cite{sano-furukawaChangeCompressibilityDAlOOH2009}.

In the 20–26~\unit{\GPaQHA} stage (or $\sim$12~\unit{\GPaEXP} onwards), \textit{mc}-QHA predicts HOC-22 and -11 configurations determine the overall compressibility. These configurations have larger volumes. The $\sim$0.4~\unit{\cubic\angstrom} discontinuous volume increases at 20 and 26~\unit{\GPaQHA} (see Fig.\,\ref{fig:4}(c)) should somehow be smoothed in the QHA compression curve leading to a further decrease of compressibility, as it seems to occur at 14~\unit{\GPaEXP} in experiments \cite{sano-furukawaChangeCompressibilityDAlOOH2009}. More advanced quantum simulations are desirable to shed light on proton states and configurations in this pressure range. This will be discussed in greater detail when the H-bond’s compressional behavior is addressed in Sec.\,\ref{sec:3D}.

Fig.\,\ref{fig:4}(c) also compares \textit{mc}-QHA compression curves of AlOOH and AlOOD at 300~K with measurements. The deuteration of \textgreek{δ} contributes to a $\sim$0.1–0.2~\unit{\cubic\angstrom} volume increase in the 0–12~\unit{\GPaQHA} range. This anomalous volume isotope effect \cite{umemotoNatureVolumeIsotope2015} is in excellent agreement with measurements \cite{sano-furukawaChangeCompressibilityDAlOOH2009, sano-furukawaDirectObservationSymmetrization2018}.

Although PBE calculations systematically overestimate pressure by $\sim$4–6~GPa or volume by $\sim$2~\unit{\cubic\angstrom}, our overall segmented \textit{mc}-compression curve shape in Fig.\,\ref{fig:4}(c) seems consistent with the experimental compression curve \cite{sano-furukawaChangeCompressibilityDAlOOH2009, sano-furukawaDirectObservationSymmetrization2018, kuribayashiObservationPressureinducedPhase2014}, particularly in the 0–14~\unit{\GPaQHA} range. Because protons should be tunneling at and beyond $\sim$18~\unit{\GPaQHA} (12~\unit{\GPaEXP}), our results are not predictive beyond this pressure. However, the phase consisting of HOC-11*, HOC-11, and HOC-22 components above $\sim$15~\unit{\GPaQHA} displays an overall compression behavior in good agreement with experiments.

\subsubsection*{Implications for the (001)-projected proton distribution}

As indicated in Sec.\,\ref{sec:3A}, unconstrained disorder is unlikely in individual $a, b$-planes, at least in short- to medium-range scale. Disorder is thus restricted to the $c$-direction. Previously measured proton distribution maps vs.\ pressure (Fig.\,\ref{fig:3} in \cite{sano-furukawaDirectObservationSymmetrization2018}) correspond to projections of the proton distribution on (001). Here we examine the correlation between the (001)-projection of the proton arrangements to the state of disorder along the $c$-direction.

Consider single crystals of each of the four HOC structures in Fig.\,\ref{fig:1}. The HOC-11’s and -22’s (001)-projections will show a single proton distribution peak. HOC-11*’s will show two symmetric proton distribution peaks; HOC-12’s will have two asymmetric peaks. Next, consider entirely random stackings of equivalent configurations for each of these HOC units along [001]. The (001)-projected proton distributions will show two symmetric proton distribution peaks for any of these randomly stacked configurations. These are, of course, the two extreme cases. The coexistence of multiple configurations would introduce an additional degree of disorder. Therefore, asymmetric peaks in the (001)-projected proton distribution map require, at least to some extent, preservation of some order along [001]. 

“Partial disorder” as measured at 6.37~\unit{\GPaEXP} (see Fig.\,\ref{fig:3}(a) in \cite{sano-furukawaDirectObservationSymmetrization2018}) best describes the proton distribution at low pressures (i.e., $< 8$ \unit{\GPaEXP}). The proton sites in the 6.37~GPa distribution map are discrete, showing H-bonds are well-defined. Some ordering in the predominant configuration, i.e., HOC-12, could also explain the asymmetric peak distribution. The continuous population change from HOC-11 to -11* and the increasing disordering trend could reduce peak asymmetry \cite{sano-furukawaDirectObservationSymmetrization2018} under compression. A recent low-field $^1$H-NMR measurement at 5.6~\unit{\GPaEXP} \cite{trybelAbsenceProtonTunneling2021} confirms that H-bond tunneling does not occur in this pressure range, leaving H-bond disorder the predominant effect at low pressure.

Around 8.4~\unit{\GPaEXP} and below 9.5~\unit{\GPaEXP} (Fig.\,\ref{fig:3}(b) in \cite{sano-furukawaDirectObservationSymmetrization2018}), the proton distribution peaks show broad and asymmetric distribution \cite{sano-furukawaDirectObservationSymmetrization2018}. This behavior signifies that protons are no longer localized at one site but can move through/across the energy barrier, which should be low (see Sec.\,\ref{sec:3A}) and have a more gentle curvature. Based on the pressure of this anomaly and the disappearance of OH-stretching band in 300~K QTB MD \cite{bronsteinThermalNuclearQuantum2017} at corresponding pressure, we correlate this behavior to HOC-12’s instability (depicted in Fig.\,\ref{fig:3}(g)) and proton’s tunneling-like behavior in a narrow pressure range preceding this instability (Sec.\,\ref{sec:3A}). Therefore, the 8.4~\unit{\GPaEXP} (001)-projected distribution map seems to have captured an intermediate state which serves as pathway for the HOC-12 to HOC-11* (aligned to half-aligned) configuration change.

A bimodal distribution with the same peak height at slightly higher pressure is observed at 9.5~\unit{\GPaEXP} \cite{sano-furukawaDirectObservationSymmetrization2018}. This behavior is consistent with the predominance of the HOC-11* configuration starting at $\sim$11~\unit{\GPaQHA} as indicated by the \textit{mc}-supercell calculation. It is also compatible with the absence of a stretching band in the QTB MD simulation \cite{bronsteinThermalNuclearQuantum2017}.

The current study cannot fully address \textgreek{δ}’s state after HOC-11* shows signs of tunneling and instability (starting at $\sim$18–20~\unit{\GPaQHA} or $\sim$12–14~\unit{\GPaEXP}). The increase in OH bond-length (next section) and the lower compressibility after -11*’s instability seem to suggest that a transformation back into HOC-11 and -22, the proton-aligned configurations, is possible. However, the existence of these configurations is inconsistent with the absence of OH-stretching band frequencies in 300~K QTB MD \cite{bronsteinThermalNuclearQuantum2017}, which suggests an everlasting dominance of the half-aligned configuration up to the H-bond symmetrization pressure. This behavior could be caused by strong anharmonic effects and tunneling not included in our study. Nevertheless, the dominance of the half-aligned configuration (HOC-11*), disordered mixtures of aligned ones (HOC-11 and -22), or a dynamically disordered \textgreek{δ} could all contribute to the (001)-projected proton distribution showing two symmetric peaks before full H-bond symmetrization.

Suppose our theory represents approximately the behavior of proton configurations under pressure. Our results agree that \textgreek{δ} is partially disordered at low pressures \cite{sano-furukawaDirectObservationSymmetrization2018, kuribayashiObservationPressureinducedPhase2014}. However, our results suggest that the previously proposed “fully-disordered” stage \cite{sano-furukawaDirectObservationSymmetrization2018, kuribayashiObservationPressureinducedPhase2014} might, in fact, be a predominantly half-aligned HOC-11*-type configuration starting at around 8.5~\unit{\GPaEXP}.

\subsection{Evolution of neutron diffraction intensity and the 9~GPa transition}
\label{sec:3D}

\begin{figure*}
    \centering
    \includegraphics[width=.45\textwidth]{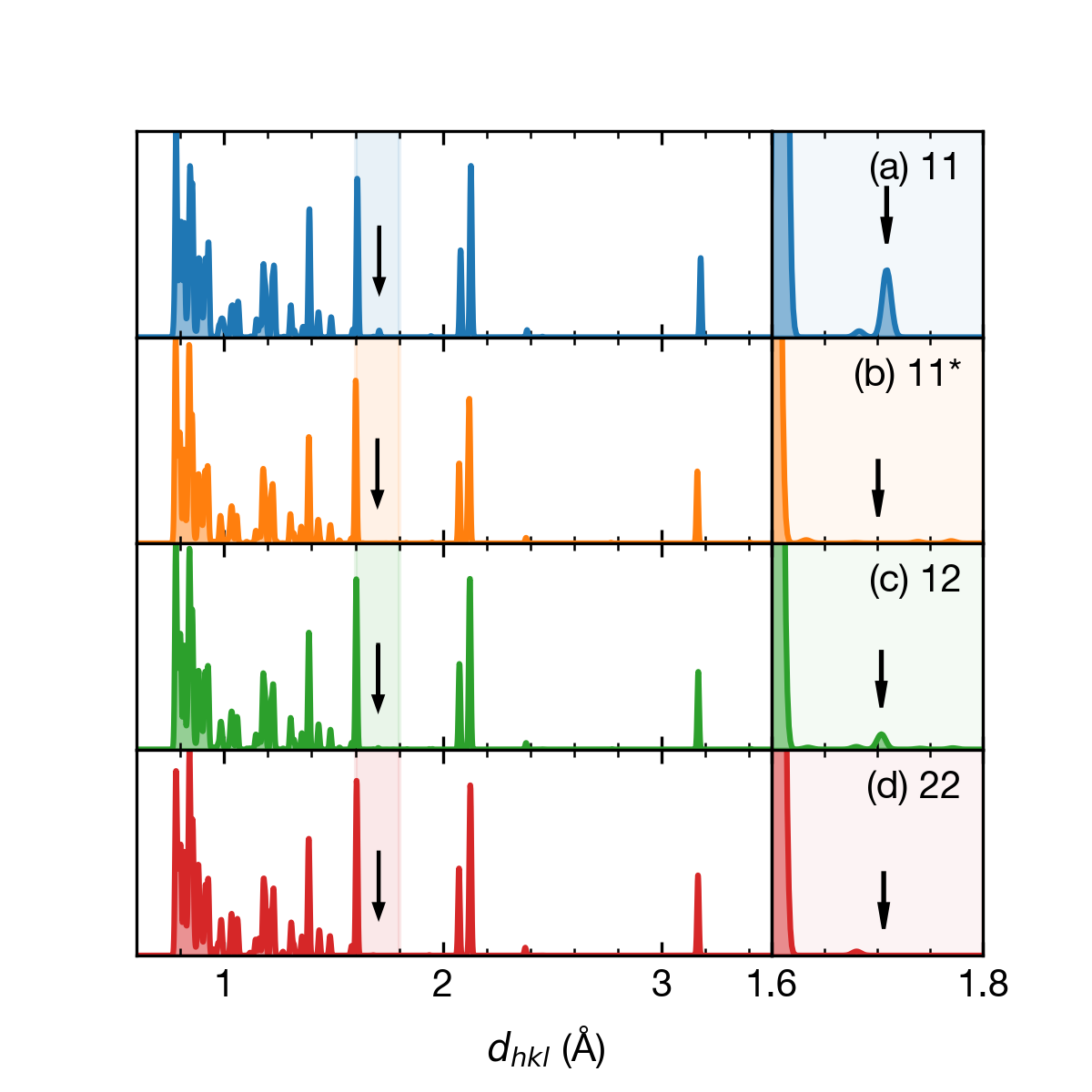}
    \includegraphics[width=.45\textwidth]{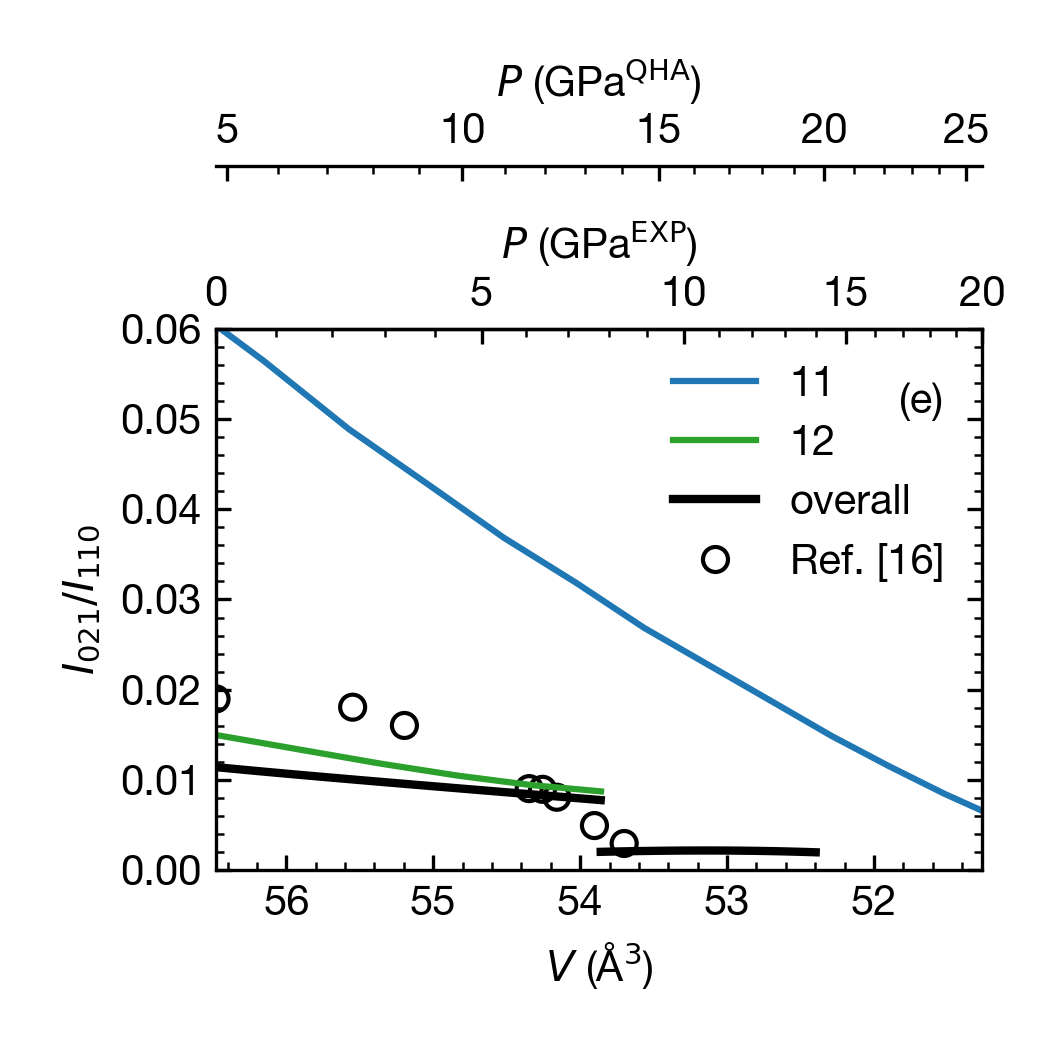}
    \caption{(a–d) Diffraction pattern at $d_{hkl} = 0.6\text{--}3.5$~Å for 4 HOC supercell structures at 0~\unit{\GPaPBE}. The diffraction intensity is normalized to 1. The shaded panel is a zoom-in view at the 021 peak region ($d_{hkl} = 1.6\text{--}1.8$~Å, intensity amplified 10x). The arrow points to the expected $d_{hkl}$ position ($\sim$1.7~Å) of the 021 peak. (e) Relative diffraction intensities of 021 peaks normalized by that of the 110 peak for HOC-11, -22 (colored curves), and “overall” (bold black solid curves) vs.\ volume compared. Open circles are data reported in Ref.\,\cite{sano-furukawaDirectObservationSymmetrization2018}. The relative 021 peak intensities for HOC-11* and -22 are less than 0.002 and, therefore, not plotted. The “overall” peak intensity is the average of peak intensities from different configurations using the \textit{mc}-QHA populations at 300~K. Diffraction patterns, peak positions, and peak intensities were calculated using CrystalDiffract® \cite{CrystalDiffractPowderDiffraction} and the optimized crystal structures.}
    \label{fig:5}
\end{figure*}

Previous studies \cite{sano-furukawaDirectObservationSymmetrization2018, kuribayashiObservationPressureinducedPhase2014} constrain the pressure of the partially disordered to fully disordered transition to $\sim$9~\unit{\GPaEXP} by tracking the pressure dependence of the 021 neutron diffraction peak intensity. Because this peak exists in the “partially-disordered” model and not in the “fully-disordered” model, the evolution of the 021 peak intensity should track the state of disorder. However, the configuration population analysis in Sec.\,\ref{sec:3C} suggested the 9~\unit{\GPaEXP} configuration change could be better explained by the disappearance of HOC-12-like proton configurations and its transformation into the HOC-11*-like configurations (“HOC-like configurations” means domains, large or small, of these ordered configurations). For consistency, we need to demonstrate a change from partially disordered HOC-configurations to half-aligned configurations could also result in the same peak extinction behavior.

A structure factor analysis can determine the extinction condition for the 021 diffraction peak ($F_{021}$, or $F_{022}$ for the $1 \times 1 \times 2$ supercell \footnote{In $1 \times 1 \times 2$ supercells, the Miller indices for the peak corresponding to experimentally observed 021-peak should be 022, but for convenience and without confusion, we continue to call it 021 henceforth. The Miller indices of the 110-peak used for normalization do not change for $1 \times 1 \times 2$ supercell.}) of the four HOC configurations. Results show that $F_{021} \neq 0$ for HOC-11 (primitive cell space group $P2_1nm$, No.\,31) and HOC-12 (supercell space group $Pm$, No.\,31). $F_{021}=0$ for HOC-22 (primitive-cell space group $Pn21m$, No.\,31) and HOC-11* (supercell space group $P2_12_12_1$, No.\,19). In the latter structures, the corresponding peak is absent. Table S1 gives detailed information on proton site symmetries in these structures. With the help of CrystalDiffract® \cite{CrystalDiffractPowderDiffraction}, we create the diffraction profiles for \textit{ab initio} optimized HOC structures and get $d$-spacings ($d_{hkl}$) of the diffraction peaks.

Fig.\,\ref{fig:5}(a–d) shows the normalized diffraction patterns of the four supercells at 0~\unit{\GPaPBE}. Most features in our diffraction spectra agree with those in neutron diffraction patterns \cite{sano-furukawaDirectObservationSymmetrization2018}. An observed systematic trend is that low $d$-spacing peaks have low diffraction intensities \cite{sano-furukawaDirectObservationSymmetrization2018}. This trend is not present in our calculated diffraction patterns, but it should be irrelevant to our analysis focused on the pressure evolution of one particular peak intensity. The calculated $d_{021}$ spacing is $\sim$1.7~Å, consistent with that reported \cite{sano-furukawaDirectObservationSymmetrization2018, sano-furukawaNeutronDiffractionStudy2008}. In Fig.\,\ref{fig:5}(a–d), $d_{021}$’s are indicated with arrows. $d_{021}$’s differ slightly for different HOC structures at the same PBE pressure because lattice parameters differ slightly. CrystalDiffract® \cite{CrystalDiffractPowderDiffraction}’s prediction shows that the 021 peaks exist only in HOC-11 and -12 but not in -11* and -22, consistent with our structure factor analysis. Therefore, transformations from HOC-11 or -12 configurations into HOC-11* will extinguish the 021 diffraction peak.

To further illustrate our analysis and to compare it with the reported compressional behavior of the 021 peak \cite{sano-furukawaDirectObservationSymmetrization2018}, we plot the peak intensities ratio, $I_{021}/I_{110}$, vs.\ volume in Fig.\,\ref{fig:5}(e). Our results show $I_{110}$’s are independent from H-bond configurations, therefore it is reasonable to use it as the standard for normalization. The $I_{021}/I_{110}$ ratio is smaller than 0.002 for HOC-11* and -22 over the entire pressure range, thus not shown in Fig.\,\ref{fig:5}(e). The $I_{021}/I_{110}$ ratios for both HOC-11 and -12 decay linearly under compression, but neither is absent at $\sim$8~\unit{\GPaEXP}. For this reason, we confirm that a configuration change, continuous or discrete, from HOC-11 or -12 is likely at 9~\unit{\GPaEXP}. As shown in Sec.\,\ref{sec:3C}, it is the HOC-12 configuration that disappears at this pressure. The HOC-11 population decreases in favor of the HOC-11*’s up to 9~\unit{\GPaEXP}, but it survives beyond this pressure and is the last one to disappear in our calculations.

Considering the contribution of each configuration based on their 300~K populations (see Fig.\,\ref{fig:4}(a)), a curve reflecting our estimated population-averaged relative intensity ratio $I_{021}/I_{110}$ is also plotted in Fig.\,\ref{fig:5}(e). Compared with Ref.\,\cite{sano-furukawaDirectObservationSymmetrization2018}, our estimation of $I_{021}/I_{110}$ in the 0–9~\unit{\GPaEXP} range is slightly smaller than the experimental measurement, but the pressure dependence of this average $I_{021}/I_{110}$ shows the correct decaying trend up to 9~\unit{\GPaEXP}. The disappearance of HOC-12 at $\sim$9~\unit{\GPaEXP} and the dominance of HOC-11* afterwards makes $I_{021}/I_{110}$ decay to below 0.002 or become effectively absent above 9~\unit{\GPaEXP}, in agreement with Ref.\,\cite{sano-furukawaDirectObservationSymmetrization2018}’s observation. The decay and extinction of $I_{021}/I_{110}$ predicted by the \textit{mc}-QHA calculations is clearly associated with changes in configuration populations and the disappearance of the HOC-12 configuration at $\sim$9~\unit{\GPaEXP} ($\sim$13~\unit{\GPaQHA}). Therefore, it supports the partially disordered to predominantly HOC-11*-ordered proton configuration change we described in the Sec.\,\ref{sec:3C}. However, the present study is unable to describe the exact configuration change behavior near the vibrational stability limit, therefore only a sharp transition is shown in Fig.\,\ref{fig:5}(e) instead of more gradual $I_{021}/I_{110}$ ratio decay as measurements \cite{sano-furukawaDirectObservationSymmetrization2018, sano-furukawaNeutronDiffractionStudy2008}. As described in Sec.\,\ref{sec:3C}, the transition from HOC-12 to -11* should undergo a brief regime of H-bond tunneling, where protons have a large position spread \cite{sano-furukawaDirectObservationSymmetrization2018}, weakening the $I_{021}$ diffraction intensity as a result. At this point, tunneling might not have developed to its full-fledged behavior before HOC-12 transforms into -11*. Tunneling here should be viewed as a pathway for change in H-bond arrangement only. Further experimental and theoretical investigations are needed to detail this proton configuration change phenomenon at 9~\unit{\GPaEXP}. In addition to other evidence presented earlier, the good agreement between the predicted and measured development of the $I_{021}$ diffraction peak intensity provides further evidence that the observed $\sim$9~\unit{\GPaEXP} change could correspond primarily to a change from predominant HOC-12 to -11* configuration.

After a brief regime of HOC-11* stability, \textit{mc}-supercell calculations predict HOC-11* transforms into HOC-22 and -11 before full H-bond symmetrization. If so, the reappearance of the 021 peak might be expected alongside -11’s reappearance at $\sim$26~\unit{\GPaQHA} ($\sim$18~\unit{\GPaEXP}). This effect has not been reported. Suppose tunneling or anharmonic fluctuations are happening between HOC-22 and -11. In that case, H-bond may appear symmetric, as implied by merging the two symmetric H-distribution peaks in the 18.1~\unit{\GPaEXP} (001)-projection map \cite{sano-furukawaDirectObservationSymmetrization2018}. This behavior is actually suggested by the 300~K QTB MD showing a single high-frequency H-bond stretching bandwidth \cite{bronsteinThermalNuclearQuantum2017}. In this case, the 021 neutron peak is not expected to reappear.

\subsection{Interatomic distances \textit{d}(OH) and \textit{d}(OO) and tunneling}
\label{sec:3E}

\begin{figure}
    \centering
    \includegraphics[width=.5\textwidth]{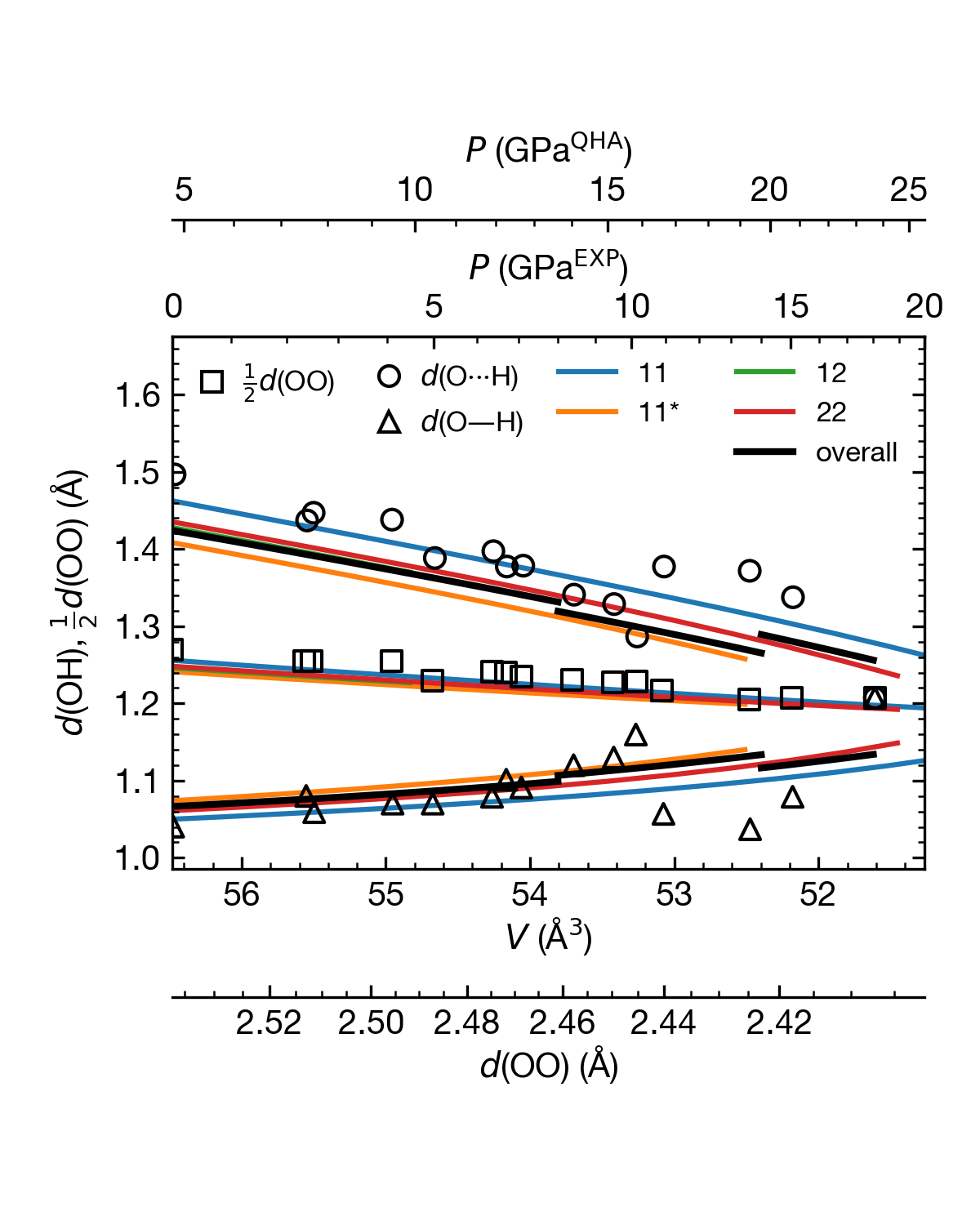}
    \caption{Interatomic distances $d$(OO) and $d$(OH) vs.\ unit-cell volume (2~f.u.) for four \textgreek{δ}-AlOOH configurations indicated by color. $d$(OH)s larger and smaller than 1.2~Å correspond to H-bonds and ionic bonds, respectively. Black lines denote $d$(OH)’s averaged by population ($n_m$). Scattered symbols denote $d$(OO)’s and $d$(OH)’s reported in Ref.\,\cite{sano-furukawaDirectObservationSymmetrization2018}. Additional axes showing 300~K experimental pressure, 300~K \textit{mc}-QHA pressure, and the experimental $d$(OO) at corresponding volumes are also offered to facilitate comparison. Volume in the bottom axis corresponds to 2~f.u..}
    \label{fig:6}
\end{figure}

This section investigates the evolution of interatomic distances $d$(OO) and $d$(OH) under pressure. Fig.\,\ref{fig:6} shows the volume dependence of $d$(OO) and $d$(OH) bond-lengths. We first notice that $d$(OO) depends almost linearly on volume in the 0–20~\unit{\GPaEXP} pressure range. Whether measured \cite{sano-furukawaDirectObservationSymmetrization2018} or computed, this bond-length is undisrupted either by protons’ anomalous behavior \cite{sano-furukawaDirectObservationSymmetrization2018} or by the differences between H-bond arrangements in \textit{ab initio} calculations. Therefore, plotting $d$(OH) vs.\ volume or vs.\ $d$(OO) in Fig.\,\ref{fig:6} is equally adequate to relate protons’ local environment. We provide additional scales indicating the experimental $d$(OO) and the experimental and 300~K \textit{mc}-QHA pressures for reference in Fig.\,\ref{fig:6}. Our \textit{ab initio} $d$(OO) values agree well with experimental ones \cite{sano-furukawaDirectObservationSymmetrization2018} despite a slight but systematic underestimation of $\sim$0.02~Å.

Fig.\,\ref{fig:6} also shows measured and calculated $d$(O—H) and $d$(O···H) for each of the four HOC configurations as well as for the “average” bond lengths weighted by the HOC populations. In the 0–11~\unit{\GPaEXP} pressure range, the measured $d$(O—H) fluctuates within the uncertainty range defined by the four HOC configurations, and it is well-predicted in general. The measured $d$(O···H) at low pressures is slightly larger than our \textit{ab initio} values. The more accurate prediction of $d$(O—H) than $d$(O···H) is a general feature of standard DFT functionals, i.e., GGA-PBE in this case. Similar effects have been observed in previous studies of ice (see, e.g., \cite{umemotoOrderDisorderPhase2010}). The $d$(O···H) and $d$(O—H) values, both measured \cite{sano-furukawaDirectObservationSymmetrization2018} and calculated averages up to 8.5~\unit{\GPaEXP}, are generally bracketed by HOC-11 and -11* calculated values. Under compression, measured $d$(O···H) and $d$(O—H) \cite{sano-furukawaDirectObservationSymmetrization2018} change more rapidly with volume in the 0–11~\unit{\GPaEXP} range than the calculated ones for any of the four HOC configurations. The expected continuous depopulation of HOC-11 in favor of -11* in the 0–8~\unit{\GPaEXP} range and the configuration change from HOC-12 to -11* at $\sim$8~\unit{\GPaEXP} could reasonably explain this trend since both changes contribute to the accelerated decrease in $d$(O···H) and increase in $d$(O—H). 

At $\sim$11~\unit{\GPaEXP}, $d$(OO), $d$(O—H), and $d$(O···H) in HOC-11* approach $\approx$2.44~Å, $\approx$1.14~Å, and $\approx$1.29~Å respectively, as shown in Fig.\,\ref{fig:6}. \textit{mc}-supercell modeling (bold “average” curve in Fig.\,\ref{fig:6}) predicts an abrupt increase in $d$(O···H) and decrease in $d$(O—H) due to the change from HOC-11* into -22 at $\sim$20~\unit{\GPaQHA}. Meanwhile, at similar critical $d$(OO) and $d$(OH), the evolution of measured $d$(OH) vs.\ pressure also becomes unconventional. Structure refinement \cite{sano-furukawaDirectObservationSymmetrization2018} shows that: a) $d$(O···H) and $d$(O—H) abruptly decrease and increase, respectively near 11~\unit{\GPaEXP}; b) an abrupt increase and decrease in $d$(O···H) and $d$(O—H) then follows; c) in the 11–18~\unit{\GPaEXP} pressure range, $d$(O···H) and $d$(O—H) reestablish the decreasing and increasing behaviors and gently continue with this trend until they finally equalize at 18~\unit{\GPaEXP}, indicating H-bond symmetrization.

Between 9.5–11.5~\unit{\GPaEXP} ($\sim$15–18~\unit{\GPaQHA}), , protons are not tunneling because the 9.5~\unit{\GPaEXP} measured (001)-projected proton distribution shows a well-defined double-peak shape \cite{sano-furukawaDirectObservationSymmetrization2018} and HOC-11* remains stable ($\sim$12–20~\unit{\GPaQHA}). \textit{Ab initio} calculations should predict well the predominant H-bond configuration in this volume regime. The decreasing $d$(O···H) and increasing trend in $d$(O—H) do not exceed their respective limits ($d$(O—H) $\approx$ 1.14~Å, and $d$(O···H) $\approx$ 1.29~Å). Further compression causes an anharmonic effect (mixing of OH stretching and bending modes, see Sec.\,\ref{sec:3B} and Figs.\,\ref{fig:3}(b,f,j)), structural instability in \textit{ab initio} calculations, and discontinuities in measured bond-length trends \cite{sano-furukawaDirectObservationSymmetrization2018}. Therefore, the $\sim$11~\unit{\GPaEXP} discontinuities in measured bond-lengths \cite{sano-furukawaDirectObservationSymmetrization2018} and the \textit{ab initio} instability in HOC-11*, possibly accompanied by a pre-instability tunneling behavior, are likely correlated. If so, our static \textit{ab initio} calculation predicts well the $d$(OO) and both $d$(OH)s limiting values. This $d$(O—H) limit is reflected on the measured OH-bending mode frequency \cite{kagiInfraredAbsorptionSpectra2010} vs.\ volume. This limit separates two regimes distinguished by this mode frequency’s pressure/volume dependence (see Fig.\,S4(a)). In \textit{ab initio} phonon calculations, OH-stretching and in-plane bending bands start mixing at this point (see Sec.\,\ref{sec:3B}, Figs.\,\ref{fig:3}(e–h)).

\subsection{Implication}

Our work shows that the H-bond disordering process that precedes H-bond symmetrization can be described as a sequence of changes in predominant H-bond configurations. Subtle changes in the atomic structure under pressure, as found here, contribute to anelastic effects. We identify such an effect in the pressure range of $\sim$10–20~\unit{\GPaQHA}, which corresponds approximately to the pressure range of $\sim$6–15~\unit{\GPaEXP}. In the latter pressure range, Brillouin scattering experiments identify a rapid increase in acoustic velocities and anomalies in the compression curve. Here we suggest that this rapid increase in acoustic velocities is related to anelastic effects, i.e., \textit{structural accommodation} under pressure, before H-bond symmetrization. Anelastic effects are known to cause seismic wave attenuation. As previously suggested \cite{mashinoSoundVelocitiesDAlOOH2016}, fast velocity anomalies in the transition zone (15–25~\unit{\GPaEXP}) may signal the presence of \textgreek{δ} in this region. Here we suggest that if seismic attenuation is observed concurrently with fast seismic velocities (low-frequency acoustic waves) in the transition zone (410–660~km depth), it will strengthen the evidence for the presence of \textgreek{δ} in this region. High crust/slab temperatures in Earth’s interior might further enhance H-bond disordering, and stiffening in compressibility will likely occur at even lower pressures. The magnitude of these effects in the mantle depends on the relative abundance of \textgreek{δ}, which correlates with the degree of slab hydration and is quite uncertain. 

These predicted anelastic effects are expected in other structurally similar water carriers in the mantle, e.g., \ce{MgSiO4H2} phase H \cite{tsuchiyaFirstPrinciplesPrediction2013}, \ce{\epsilon-FeOOH} \cite{xuSolubilityBehaviorDAlOOH2019}, and Al,H-bearing stishovite. They might also be expected in other dissimilar phases since the phenomena addressed in this paper are typical of H-bonds. The importance of these effects to seismic velocities will be proportional to the abundance of the hydrous phase in the hydrated slab, which is uncertain \cite{ohtaniHydrousMineralsStorage2015}. Therefore, a precise characterization of \textgreek{δ}-AlOOH’s acoustic velocities and other co-existing hydrous phases will also help clarify the degree of slab hydration in the mantle up to CMB depths.

At 300~K, our calculations are not predictive beyond 18~\unit{\GPaQHA} ($\sim$12~\unit{\GPaEXP}), where anharmonic effects become strong. Therefore, it is useless to speculate further on their geophysical consequences. Predictive calculations beyond this pressure will require genuinely quantum MD simulations to address H-bonds at high temperatures, encompassing the superionic and the dehydration regimes \cite{houSuperionicIronOxide2021, duanPhaseStabilityThermal2018} that occur near ambient mantle temperatures.

\section{Conclusion}
\label{sec:conclusion}

In summary, using multi-configuration quasiharmonic (\textit{mc}-QHA) \textit{ab initio} calculations, we studied a sequence of H-bond configuration changes in H-off-center \textgreek{δ}-AlOOH (HOC-\textgreek{δ}) under pressure at 300~K, up to full H-bond symmetrization. 

The energy barrier for proton jump calculations confirms the notion that H-bond disordering is limited by the presence of one \ce{OH-} per aluminum octahedron. This constraint imposes short to medium range H-bond ordering in the $a, b$-plane and limits disorder to the $c$-direction. Short-range interaction between $a, b$-planes with different H-bond arrangements allows us to use a $1 \times 1 \times 2$ supercells to model the disordered \textgreek{δ} phase. Therefore, we are able to use four symmetrically inequivalent supercell configurations (16 configurations in total), HOC-11, -12, -11*, and -22, to investigate the change in configuration population vs.\ pressure with \textit{mc}-QHA. HOC-11 and -22 contain protons fully aligned in the [001] interstitial channels, HOC-11* has only half-aligned protons, while HOC-12 has both proton arrangements (see Fig.\,\ref{fig:1}). This model consistently reproduces a significant number of experimental observations and allows us to correlate some measured properties with underlying atomistic processes. We do this by inspecting calculated and measured properties vs.\ volume rather than pressure.

Below 8~\unit{\GPaEXP}, there is considerable variation in configuration populations (see Fig.\,\ref{fig:4}(a)), but HOC-12, with two kinds of proton alignments, is the most abundant one. \textit{mc}-QHA calculations reproduce well both $d$(O···H) and $d$(O—H) bond lengths (see Fig.\,\ref{fig:6}) and \textgreek{δ}’s more compressive behavior in this pressure range \cite{sano-furukawaChangeCompressibilityDAlOOH2009} (see Figs.\,\ref{fig:4}(b,c) and S4(d)). They result from the continuous change in H-bond configuration populations under pressure. Also, the coexistence of multiple configurations (\textit{mc}) confirms \textgreek{δ}’s partially disordered nature, which should result in the asymmetric proton distribution projected in (001) \cite{sano-furukawaDirectObservationSymmetrization2018}. \textit{mc} coexistence also results in the four broad peaks observed in 0~GPa Raman spectra \cite{tsuchiyaVibrationalPropertiesDAlOOH2008, ohtaniStabilityFieldNew2001} and the two OH-stretching bands observed in 300~K QTB-MD simulations \cite{bronsteinThermalNuclearQuantum2017}, one from aligned and the other from half-aligned proton configurations. 

At $\sim$8~\unit{\GPaEXP}, the 021 neutron diffraction peak disappears \cite{sano-furukawaDirectObservationSymmetrization2018} (see Fig.\,\ref{fig:5}(e)), the proton distribution peaks become ill-defined \cite{sano-furukawaDirectObservationSymmetrization2018}, and the pressure dependence of the out-of-plane OH-bending mode frequency changes \cite{kagiInfraredAbsorptionSpectra2010} (see Fig.\,S4(a)). Also, the higher OH-stretching frequency band disappears in the 300~K QTB-MD simulations around $\sim$10~GPa \cite{bronsteinThermalNuclearQuantum2017}. We correlate these observations with the vibrational instability and disappearance of the predominant HOC-12 configuration (see Figs.\,\ref{fig:3}(c,g,k), \ref{fig:4}(a), \ref{fig:5}(e), and S4(a)) and its replacement by -11* with only half-aligned protons. The obscure proton distribution peaks might result from a brief proton tunneling behavior preceding this configuration change.

Between $\sim$8 and $\sim$11.5~\unit{\GPaEXP}, proton distribution maps show two symmetric discrete peaks that have been attributed to a fully disordered proton configuration \cite{sano-furukawaDirectObservationSymmetrization2018}, while 300~K QTB-MD simulations display a single OH-stretching frequency band \cite{bronsteinThermalNuclearQuantum2017}. Overall, \textgreek{δ} becomes less compressible in this brief pressure range \cite{mashinoSoundVelocitiesDAlOOH2016, sano-furukawaChangeCompressibilityDAlOOH2009} (see Figs.\,\ref{fig:4}(c) and S4(d)). $d$(O—H) and $d$(O···H) reach their largest and smallest values, respectively, at $\sim$11.5~\unit{\GPaEXP} (see Fig.\,\ref{fig:6}). We correlate these observations with the predominance of HOC-11*, small abundances of -11 and -22 in this pressure range, and the eminent instability of HOC-11* at $\sim$11.5~\unit{\GPaEXP}. HOC-11*, the densest configuration (see Fig.\,\ref{fig:4}(b)), produces two discrete symmetric peaks (see Fig.\,\ref{fig:1}), is less compressible, and displays smaller compressibility, and a single and lower OH-stretching frequency band (compare Figs.\,\ref{fig:3}(b) and \ref{fig:3}(c)). According to our \textit{mc}-QHA calculation, the proton configuration is “more ordered” rather than “more disordered” in this pressure range corresponding to 11–18~\unit{\GPaQHA}. The absence of anharmonic and tunneling effects in our calculations and the problematic DFT description of H-bonds seem to constrain unrealistically the pressure stability field of HOC-11* up to 11.5~\unit{\GPaEXP}. Experimental observations beyond this pressure correlate at best qualitatively with our \textit{mc}-QHA results at higher nominal pressures.

In the $\sim$11.5–18~\unit{\GPaEXP} pressure range, $d$(O···H) and $d$(O—H) bond lengths behave anomalously (see Fig.\,\ref{fig:6}). Their pressure dependences are abruptly reversed after reaching limiting values at $\sim$11.5~\unit{\GPaEXP}. However, they resume their normal pressure-induced behavior, i.e., decrease in $d$(O···H) and increase in $d$(O—H) with pressure soon after. OH-bending mode frequencies also have different compressive behavior below and above $\sim$11.5~\unit{\GPaEXP} \cite{sano-furukawaChangeCompressibilityDAlOOH2009} (see Fig.\,S4(a)). These changes suggest a proton configuration change at this pressure followed by normal but accelerated bond length compressive behavior, terminating in H-bond symmetrization at 18~\unit{\GPaEXP}. Mixing of OH-bending and stretching modes in HOC-11* (anharmonicity) starts at $\sim$18~\unit{\GPaQHA} ($\sim$11.5~\unit{\GPaEXP}) (see Sec.\,\ref{sec:3B}, Figs.\,\ref{fig:3}(b,f,j)), indicating this latter anomalous stage involves anharmonic effects and is likely accompanied by tunneling. Although tunneling is anticipated close to H-bond symmetrization, it is also expected near configuration changes. HOC-11* becomes unstable at $\sim$20~\unit{\GPaQHA} ($\sim$14~\unit{\GPaEXP}), and only HOC-22 and -11 configurations, both with fully aligned protons and with the same multiplicity (see Fig.\,\ref{fig:1}) survive after that (see Fig.\,\ref{fig:4}(a)). Both configurations have slightly larger volumes than HOC-11* and might explain the decrease in compressibility in this pressure range (see Figs.\,\ref{fig:4}(b,c)). Suppose these configurations are present in this pressure range. In that case, \textit{mc}-QHA results predict the reappearance of the highest OH-stretching frequency modes or rather an increase of the OH-stretching band frequency (compare Fig.\,\ref{fig:3}(b) with Figs.\,\ref{fig:1}(a,d)), which has not been reported yet. Coherent proton tunneling within some length scale between these configurations is consistent with pre-H-bond symmetrization behavior. 

A summary of experimental observations and calculated results are presented in Table~\ref{tab:summary}. Large scale path integral MD calculations using more predictive exchange-correlation functionals are desirable to elucidate further the relationships between collective proton behavior predicted by \textit{mc}-QHA calculations and experimental observations.

\begin{table*}[]
    \centering
    \caption{Summary of 300~K phase changes in \textgreek{δ}}
    \raggedright
    
    \renewcommand{\arraystretch}{1.35}
    
    \begin{ruledtabular}
    \begin{tabular}{p{0.18\textwidth} p{0.30\textwidth} p{0.50\textwidth}}
    
        Pressure &
        State of \textgreek{δ} &
        Experimental observations and simulation findings
        \\\hline
        \begin{minipage}[t]{0.18\textwidth}\raggedright
        0–8~\unit{\GPaEXP} \\ (0–12~\unit{\GPaQHA})
        \end{minipage} &
        
        \begin{minipage}[t]{0.30\textwidth}
        \raggedright
        Partially disordered with pressure dependent site populations
        \end{minipage} &
        
        \begin{minipage}[t]{0.50\textwidth}\raggedright
        \begin{compactitem}[\textbullet]
        \item High compressibility owing to site population change observed in \textit{mc}-QHA \cite{sano-furukawaChangeCompressibilityDAlOOH2009,mashinoSoundVelocitiesDAlOOH2016}
        \item Two OH-stretching bands observed in 300~K QTB MD \cite{bronsteinThermalNuclearQuantum2017} and in \textit{mc}-QHA
        \item Asymmetric proton distribution with two peaks of different heights \cite{sano-furukawaDirectObservationSymmetrization2018}
        \item Four broad peaks observed in 0~GPa Raman spectra \cite{ohtaniStabilityFieldNew2001} and in \textit{mc}-QHA \cite{tsuchiyaVibrationalPropertiesDAlOOH2008}
        \end{compactitem}
        \end{minipage}
        \\\hline
        \begin{minipage}[t]{0.18\textwidth}\raggedright
        $\sim$8~\unit{\GPaEXP} \\ ($\sim$12~\unit{\GPaQHA})
        \end{minipage} &
        
        \begin{minipage}[t]{0.30\textwidth}
        \raggedright
        HOC-12 changes into -11* possibly preceded by brief proton tunneling
        \end{minipage} &
        \begin{minipage}[t]{0.50\textwidth}
        \begin{compactitem}[\textbullet]
        \item Disappearance of 021 neutron diffraction peak \cite{sano-furukawaDirectObservationSymmetrization2018}
        \item Obscure peaks in proton distribution map \cite{sano-furukawaDirectObservationSymmetrization2018}
        \item First slope change in out-of-plane OH-bending mode frequency vs.\ pressure \cite{kagiInfraredAbsorptionSpectra2010}
        \item Brief tunneling accompanied by a transition from HOC-12 to HOC-11* in \textit{mc}-QHA
        \item Disappearance of the higher-frequency OH-stretching band in 300~K QTB MD \cite{bronsteinThermalNuclearQuantum2017}
        \end{compactitem}
        \end{minipage}
        \\\hline
        \begin{minipage}[t]{0.18\textwidth}\raggedright
        8–11.5~\unit{\GPaEXP} \\ ($\sim$12–18~\unit{\GPaQHA})
        \end{minipage} &
        
        \begin{minipage}[t]{0.30\textwidth}
        \raggedright
        Mostly HOC-11* (ordered)
        \end{minipage} &
        \begin{minipage}[t]{0.50\textwidth}
        \begin{compactitem}[\textbullet]
        \item Symmetric proton distribution with discrete peaks \cite{sano-furukawaDirectObservationSymmetrization2018}
        \item Increased compressibility \cite{sano-furukawaChangeCompressibilityDAlOOH2009,mashinoSoundVelocitiesDAlOOH2016}
        \item Single OH-stretching band in 300~K QTB MD \cite{bronsteinThermalNuclearQuantum2017}
        \item Predominant HOC-11* configuration in \textit{mc}-QHA also with a single OH-stretching frequency band
        \end{compactitem}
        \end{minipage}
        \\\hline
        \begin{minipage}[t]{0.18\textwidth}\raggedright
        11.5–18~\unit{\GPaEXP} \\ ($\sim$18–24~\unit{\GPaQHA})
        \end{minipage} &
        
        \begin{minipage}[t]{0.30\textwidth}
        \raggedright
        HOC-11* becomes unstable, mostly HOC-22 with some HOC-11 with possible proton tunneling
        \end{minipage} &
        
        \begin{minipage}[t]{0.50\textwidth}\raggedright
        \begin{compactitem}[\textbullet]
        \item Anomalies in OH-bond lengths \cite{sano-furukawaDirectObservationSymmetrization2018}
        \item After slope change at $\sim$11~GPa, kinks observed in in-plane OH-bending frequency \cite{kagiInfraredAbsorptionSpectra2010}
        \item OH-bending and stretching mode mixing (anharmonicity) likely accompanied by tunneling in HOC-11* starting at 11.5~\unit{\GPaEXP} (or $\sim$18~\unit{\GPaQHA})
        \item HOC-11* becomes vibrationally unstable at $\sim$20~\unit{\GPaQHA} (or $\sim$14~\unit{\GPaEXP})
        \item Unstable mode in HOC-11* expected to have a tunneling component near 20~\unit{\GPaQHA}
        \item Predicted increase of OH-stretching bond frequencies
        \end{compactitem}
        \end{minipage}
        \\\hline

        \begin{minipage}[t]{0.18\textwidth}\raggedright
        18~\unit{\GPaEXP}
        \end{minipage} &
        
        \begin{minipage}[t]{0.30\textwidth}
        \raggedright
        Symmetric H-bond
        \end{minipage} &
        
        \begin{minipage}[t]{0.50\textwidth}\raggedright
        \begin{compactitem}[\textbullet]
        \item Single OH bond length
        \item Frequency vs.\ volume slope changes again in OH-bending modes \cite{kagiInfraredAbsorptionSpectra2010}
        \end{compactitem}
        \end{minipage}
    \end{tabular}
    \end{ruledtabular}
    \label{tab:summary}
\end{table*}

\begin{acknowledgments}
RMW and CL were supported by DOE award DE-SC0019759. KU was supported by JSPS Kakenhi Grant No.\,17K05627. The authors also thank Tianqi Wan and Ziyu Cai for their assistance with calculations in the early stage of this work. Calculations were performed on the Extreme Science and Engineering Discovery Environment (XSEDE) \cite{townsXSEDEAcceleratingScientific2014}, which is supported by the National Science Foundation grant number ACI-1548562 and allocation TG-DMR180081.
\end{acknowledgments}

\bibliography{Geophysics}

\end{document}